\documentclass[seceq]{ptptex}

\newcommand{\EQ}{\begin{equation}}
\newcommand{\EN}{\end{equation}}
\newcommand{\bea}{\begin{eqnarray}}
\newcommand{\ena}{\end{eqnarray}}
\newcommand{\bdis}{\begin{displaymath}}
\newcommand{\edis}{\end{displaymath}}

\renewcommand{\a}{\alpha}
\renewcommand{\b}{\beta}
\renewcommand{\c}{\gamma}
\renewcommand{\d}{\delta}

\renewcommand{\t}{\tau}

\newcommand{\varint}{\int\!\!\!\!\!\!\! \sim}
\newcommand{\pa}{\partial}

\newcommand{\nn}{\nonumber \\}

\title{
Stochastic Gauge Fixing for ${\cal N} = 1$\\ 
 Supersymmetric 
Yang-Mills Theory}

\author{
Naohito\ Nakazawa
}

\inst{
 Theoretical Physics Laboratory, RIKEN \\
 Wako 351-0198, Japan 
}

\abst{
The gauge fixing procedure for ${\cal N}=1$ supersymmetric Yang-Mills theory (SYM) is proposed in the context of the stochastic quantization method (SQM). The stochastic gauge fixing, which was formulated by Zwanziger for Yang-Mills theory, is extended to SYM$_4$ in the superfield formalism by introducing a chiral and an anti-chiral superfield as the gauge fixing functions. It is shown that SQM with the stochastic gauge fixing reproduces the probability distribution of SYM$_4$, defined by the Faddeev-Popov prescription, in the equilibrium limit with an appropriate choice of the stochastic gauge fixing functions. We also show that the BRST symmetry of the corresponding stochastic action and the power counting argument in the superfield formalism ensure the renormalizability of SYM$_4$ in this context.
}

\begin{document}

\maketitle

\section{Introduction}

In a previous paper,\cite{Nakazawa1} we applied the stochastic quantization method (SQM)\cite{PW} to the ${\cal N}=1$ supersymmetric Yang-Mills theory (SYM). 
In 4 dimensions, the application of SQM to SYM$_4$ is formulated in terms of the superfield formalism. Based on the It${\bar {\rm o}}$ stochastic calculus, the underlying stochastic process, which is described by the superfield Langevin equation and also by the corresponding Fokker-Planck equation, preserves the global supersymmetry as well as the local gauge symmetry. 
We have postponed the detailed explanation of how SQM recovers the conventional (path-integral) probability distribution constructed by the Faddeev-Popov prescription. The main purpose of this paper is to present a proof of the equivalence of the SQM approach and the BRST invariant path-integral approach, and to demonstrate the perturbative self-consistency of the gauge fixing procedure in SQM for SYM$_4$.

In the SQM approach, as far as the expectation values of gauge invariant observables (such as Wilson loops) are concerned, it is not necessary to introduce an explicit gauge fixing procedure. 
By contrast, for gauge variant quantities, we need a gauge fixing procedure to introduce a drift force for the convergence of the longitudinal degrees of freedom. Such a procedure, the so-called stochastic gauge fixing procedure, was first formulated for Yang-Mills theory (YM),\cite{Zwanziger} and its equivalence to the Faddeev-Popov prescription in the path-integral approach was shown.\cite{BZ,HHS} It was also clarified how the extra gauge fixing term simulates the Faddeev-Popov ghost loops in the perturbative sense.\cite{Zwanziger,NOOY} The renormalization procedures in various models have been investigated in two different schemes. One is based on the so-called stochastic action.\cite{Gozzi,NY,CH,Okano} The other respects the gauge covariantly regularized Schwinger-Dyson equations.\cite{BHST} For non-supersymmetric gauge theories, in particular, renormalizability in SQM was established in terms of the Ward-Takahashi identities, which are derived from the BRST symmetry of the stochastic action.\cite{ZZ} The BRST invariance in this context has also been studied in relation to the five-dimensional\footnote{ 
There, the stochastic time $\t$ plays the role of the 5-th dimensional coordinate.
} local gauge invariance of YM$_4$,\cite{CH,KOT} \ first-class constraint systems including gravity,\cite{Nakazawa2,Nakazawa3} \ and a topological field theory approach to SQM.\cite{Baulieu-GZ} 

In this note, we formulate the stochastic gauge fixing procedure for SYM$_4$ in the superfield formalism. It is defined by introducing a chiral and an anti-chiral superfield as the gauge fixing functions. It is shown that SQM applied to SYM$_4$ with the stochastic gauge fixing procedure reproduces the standard Faddeev-Popov probability distribution under the appropriate choice of the gauge fixing functions, while the expectation values of local gauge invariant observables are independent of the gauge fixing procedure, in particular, of the specific choice of the gauge fixing functions.  The stochastic gauge fixing procedure enables us to carry out the perturbative analysis in terms of the stochastic action principle in the superfield formalism.  The power counting argument indicates the renormalizability of this approach. For SYM$_4$ in the superfield formalism, the renormalizability was investigated in the conventional BRST invariant path-integral approach.\cite{FP,GSR} We show the renormalizability of the BRST invariant stochastic action principle by applying the standard procedure in terms of the Ward-Takahashi identity for the 1-P-I vertices. 

This paper is organized as follows. In \S 2, we recapitulate the SQM approach for SYM$_4$ in terms of the superfield Langevin equation. In \S 3, we introduce the stochastic gauge fixing procedure. Section 4 presents a formal proof of the equivalence of the stochastic gauge fixing procedure and the conventional Faddeev-Popov prescription. In \S 5, we introduce the stochastic action in the BRST invariant framework. We show that there exists an extended BRST symmetry in a five-dimensional sense and that it is truncated to a nilpotent reduced BRST symmetry by integrating out auxiliary fields. In \S 6, we give the power counting argument and sketch the renormalizability of the BRST invariant stochastic action principle. Appendix A is devoted to explain the conventions to make the paper self-contained. In Appendix B, we give detailed discussion to complete the proof of the renormalizability of the BRST invariant stochastic action principle in terms of the Ward-Takahashi identities.

\section{Superfield Langevin equation for SYM$_4$}

  In the SQM approach for SYM$_4$, the superfield Langevin equation and the corresponding Fokker-Planck equation are formulated in terms of the It${\bar {\rm o}}$ stochastic calculus by exploiting the following similarity of SYM$_4$ in the superfield formalism to lattice gauge theories:\cite{Nakazawa1} 
\bea
\label{eq:SYM-LGT-correspondence}
& {} & 
{\rm dynamical\ field}  :\ 
U_\mu \in U(N) ,\ {\rm link\ variable}    \nn
& & \qquad \qquad \leftrightarrow 
U \equiv e^{2gV},\  {\rm vector\ superfield}\ V,\  V \in su(N)               \ .   \nn  
& {} & 
{\rm local\ gauge\ transformation} : 
U_\mu 
\rightarrow  
e^{i \Lambda (x)} U_\mu e^{-i\Lambda (x+\mu )}      \nn
& &  \qquad \qquad \leftrightarrow  
U 
\rightarrow 
e^{-ig \Lambda^\dagger} U e^{ig \Lambda},\       
{\rm chiral\ superfields}\ {\bar D}\Lambda = D\Lambda^\dagger = 0                       \ . \nn 
& {} & 
{\rm differential\ operator}  :\  
E_a, {\rm (left)\ Lie\ derivative}     ,\ E_a U_\mu= t_a U_\mu   \nn
& &  \qquad \qquad  \leftrightarrow  
{\hat {\cal E}}_a, \   {\rm analogue\ of\ the\ (left)\ Lie\ derivative}               \ . \nn 
& {} & 
{\rm integration\ measure}  : 
dU_\mu,\ {\rm Haar\ measure\ on}\  U(N)            \nn
& &  \qquad \qquad  \leftrightarrow 
\sqrt{G}{\cal D}V, \ {\rm analogue\ of\ the\ Haar\ measure}     \ . \nonumber
\nonumber
\ena
In lattice gauge theories, the stochastic process that describes the time evolution of link variables is well-established.\cite{DDH-GL-H-S,Nakazawa4} 
Although $U (z) \equiv e^{2gV (z)}$ is not a group element, 
the correspondence can be realized by introducing a differential operator ${\hat {\cal E}}^{(L)}_a (z)$ (  ${\hat {\cal E}}^{(R)}_a (z)$ ) and a local gauge invariant  measure $\sqrt{G}{\cal D}V$, which are analogues of the left (right) Lie derivative and the Haar measure defined on a group manifold, respectively. This correspondence is useful for the proof of the equivalence of the stochastic gauge fixing procedure and the conventional Faddeev-Popov prescription given in \S 4. It also simplifies the application of the standard procedure for the renormalizability in terms of the Ward-Takahashi identities to the BRST invariant stochastic ation principle. 

The superfield Langevin equation for $SU(N)$ SYM$_4$ is obtained by regarding $e^{2gV}$ as a fundamental variable: 
\bea
\label{eq:SYM-Langevin-eq1}
\displaystyle{\frac{1}{2g}}
( \Delta e^{2gV} )e^{-2gV} ( \t, z )
& = &
- \beta \Delta\t \displaystyle{\frac{1}{2g}} ( 
e^{2g L_V} {\cal D}{\cal W} + {\overline {\cal D}}{\overline {\cal W}} ) 
+ \Delta w  ( \t, z )         \ , \nn
\langle \Delta w_{ij} ( \t, z ) \Delta w_{mn} ( \t, z' ) 
\rangle_{\Delta w_\t}   
& = &
2 k \beta\Delta \t \Big( 
\delta_{in}\delta_{jm} - \displaystyle{\frac{1}{N}}\delta_{ij}\delta_{mn} 
\Big) \delta^8 ( z - z' )        \ .
\ena
Here,  
$\langle ... \rangle_{\Delta w_\t}$ represents that the expectation value is evaluated by means of the noise correlation at the stochastic time $\t$. $\beta$ is a scaling factor for the stochastic time which does not affect to the equilibrium distribution. The conventions employed here are explained in Appendix A. 
The Fokker-Planck equation is deduced from the Langevin equation (\ref{eq:SYM-Langevin-eq1}): 
\bea
\label{eq:SYM-Fokker-Planck-eq1}
\displaystyle{\frac{\pa}{\pa \t}} P( \t, e^{2gV} ) 
= 
\beta 
\displaystyle{\frac{4g^2}{k}}\int\!\!\!d^8z {\rm Tr} {\hat {\cal E}}^{(L)} (z) 
\Big( 
{\hat {\cal E}}^{(L)} (z) + ( {\hat {\cal E}}^{(L)} (z) S )
\Big) P( \t, e^{2gV} )           \ . 
\ena
Here, $P( \t, e^{2gV} )$ is a scalar probability density. 
Therefore, in a formal sense, 
$P( e^{2gV} ) = e^{- S}$ is a stationary solution of the Fokker-Planck equation. However, to assert the uniqueness of the equilibrium limit, we must introduce a regularization procedure to reduce the number of degrees of freedom to a finite value. It is also necessary to introduce the stochastic gauge fixing procedure for the convergence of the gauge degrees of freedom. 

The Langevin equation and the noise correlation in (\ref{eq:SYM-Langevin-eq1}) are covariant under the stochastic-time $independent$ local gauge transformation 
$e^{2gV} \rightarrow e^{-i g \Lambda^\dagger}e^{2gV}e^{ig \Lambda}$, for which $\Lambda$ and $\Lambda^\dagger$ are stochastic-time independent, provided that the noise superfield is transformed as 
\bea
\label{eq:noise-transformation}
\Delta w ( \t, z ) 
\rightarrow 
e^{-ig L_{\Lambda (z)^\dagger}} \Delta w ( \t, z )      \ . 
\ena
This means that $\Delta w$ is not a vector superfield. The introduction of the complex noise superfield $\Delta w^\dagger \neq \Delta w$ is the price we have paid for the local gauge covariance of the superfield Langevin equation (\ref{eq:SYM-Langevin-eq1}) for $e^{2gV}$. 

The superfield Langevin equation for the vector superfield $V$ is deduced from (\ref{eq:SYM-Langevin-eq1}) as 
\bea
\label{eq:SYM-Langevin-eq2}
\Delta V ( \t, z )
& = &
- \beta \Delta\t \Bigl( 
\displaystyle{\frac{L_V}{1 - e^{- 2g L_V}}} {\cal D}{\cal W} + 
\displaystyle{\frac{L_V}{e^{2g L_V} - 1}}{\overline {\cal D}}{\overline {\cal W}}       \Bigr) 
+ \Delta_w \Xi( \t, z )         \ . 
\ena
Here, $\Delta_w \Xi$ is the collective noise superfield defined as 
\bea 
\label{eq:SYM-collective-noise-eq1}
\Delta_w \Xi( \t, z )   
& \equiv & 
\displaystyle{{\frac{2gL_V}{e^{2g L_V} - 1}}} \Delta w    \nn 
& = & 
\displaystyle{\frac{g}{2}}
\Big\{ 
 {\rm coth}( gL_V ) - {\rm tanh}( gL_V ) 
\Big\} L_V ( \Delta w + \Delta w^\dagger )    \ . 
\ena
To derive the second expression, we have used the hermiticity assignment of the noise superfield (\ref{eq:noise-hermiticity-A-eq1}). 
As is clear from this expression, this collective noise is a vector superfield, i.e., $\Delta_w \Xi^\dagger = \Delta_w \Xi$. 
  Therefore, the time evolution of the vector superfield $V$ preserves its hermiticity $V^\dagger = V$. We note that the drift force term on the r.h.s. of (\ref{eq:SYM-Langevin-eq2}) is not a simple variation of the action, $\delta S/\delta V^a$, but it includes a superfield kernel $G^{ab}\delta S/\delta V^b$ for the local gauge covariance, as explained in (\ref{eq:SYM-Langevin-A-eq2}). In the weak coupling limit, our formulation is identical to those in Refs.\citen{Breit-GZ,Ishikawa,Kalivas} for the supersymmetric U(1) vector multiplet.

\section{Stochastic gauge fixing for SYM$_4$}

In the SQM approach, the stochastic gauge fixing was first introduced to solve the Gribov ambiguity\cite{Gribov} in YM$_4$, which exists in a class of covariant gauges in the conventional Faddeev-Popov prescription.\cite{Zwanziger,BZ,HHS} 
For YM$_4$, the gauge fixed Langevin equation is given by 
\bea
\label{eq:YM4-Langevin-eq1}
\Delta A^a_\mu (\t,x)
& = & - \Delta\t \Big\{ 
\displaystyle{\frac{\delta S_{\rm YM}}{\delta A^a_\mu (x)}} 
- \nabla_\mu v^a (A)
\Big\} + \Delta \eta^a_\mu (\t,x)      \  ,     \nn
\langle 
\Delta\eta^a_\mu (\t, x) \Delta\eta^b_\nu (\t, y) 
\rangle 
& = & 
2\Delta\t \delta^{ab}\delta_{\mu\nu} \delta^4 (x-y)     \ , 
\ena
where 
$ S_{\rm YM} = \displaystyle{\frac{1}{4}} \int\!\!\! d^4x F^a_{\mu\nu} F^a_{\mu\nu}$ 
and 
$
( \nabla_\mu v)^a
\equiv 
\partial_\mu v^a + f^{abc}A^b_\mu v^c 
$. $v^a (A)$ is the stochastic gauge fixing function. In the Zwanziger gauge fixing, it is chosen to be local:
\bea
\label{eq:Zwanziger-gauge-eq1}
v^a = \displaystyle{\frac{1}{\a}} (\partial\cdot A)^a         \ . 
\ena
For any functional of the gauge field, the stochastic gauge fixing term appears as the generator of the local gauge transformation in the stochastic-time evolution equation of their expectation values. Therefore, it does not affect the expectation values of local gauge invariant observables. In addition, the significant fact established in the perturbative calculations is that the extra self-interaction of the Yang-Mills field introduced by the stochastic gauge fixing term in (\ref{eq:YM4-Langevin-eq1}) simulates the Faddeev-Popov ghost contribution in the conventional approach.\cite{NOOY,Okano,BHST} 

Formally, the equivalence of the stochastic gauge fixing procedure and the Faddeev-Popov prescription is shown by specifying the stochastic gauge fixing function in such a way that the Faddeev-Popov distribution,
\bea
\label{eq:FP-equilibrium-distribution-eq1} 
P_{\rm FP} [A]
& \equiv & 
\int {\cal D}c{\cal D}{\bar c}{\cal D}B e^{- S_0}   \ ,  \nn 
S_0 
& = & 
S_{\rm YM} + \int\!\!\! d^4x \Bigl\{ 
   - \displaystyle{\frac{\a}{2}} (B^a)^2 + B^a \partial\cdot A^a 
   +  \partial{\bar c}^a \cdot \nabla c^a 
\Bigr\}       \  ,
\ena
satisfies the stationary condition for the probability distribution in equilibrium:
\bea
\label{eq:FP-equilibrium-distribution-eq2}
\int\!\!\! d^4x \displaystyle{\frac{\delta}{\delta A^a_\mu (x)}} \Bigl\{ 
\displaystyle{\frac{\delta}{\delta A^a_\mu (x)}} + 
\displaystyle{\frac{\delta S_{\rm YM}}{\delta A^a_\mu (x)}} 
- \Bigl( \nabla_\mu v[x; A]  \Bigr)^a    \Bigr\} P_{\rm FP} [A] = 0   \ .  
\ena
Here, the stochastic gauge fixing function $v^a[x; A]$ is given by  
\bea
\label{eq:FP-equilibrium-distribution-eq3}
v^a[x; A] 
& = & 
\int\!\!\! d^4y \langle c^a(x) \partial_\nu {\bar c}^b (y) \rangle    
\Bigl\{ (\nabla_\mu F_{\mu\nu})^b (y)      \nn 
& {} &      \qquad\qquad\qquad 
+ \alpha^{-1} \partial_\nu ( \partial\cdot A )^b (y) + 2g f^{bcd} \langle c^c(y)\partial_\nu {\bar c}^d (y) \rangle  \Bigr\}              \ ,
\ena
where the ghost propagator is defined by regarding $P_{\rm FP}$ in (\ref{eq:FP-equilibrium-distribution-eq1}) as the partition function.
As is clear from the expression of $v^a[x; A]$, it is non-local with respect to the gauge field, and its explicit form may be determined perturbatively with the appropriate renormalization procedure: 
\bea
\label{eq:FP-equilibrium-distribution-eq4}
v^a[x; A] 
 =  
\displaystyle{\frac{1}{\alpha}} (\partial \cdot A)^a 
+ {\cal O} (A^2)   \ .  
\ena
Although the stationary condition (\ref{eq:FP-equilibrium-distribution-eq2}) with (\ref{eq:Zwanziger-gauge-eq1}) cannot be solved for an arbitrary value of the gauge parameter $\a$, (\ref{eq:FP-equilibrium-distribution-eq4}) indicates that the Zwanziger gauge fixing (\ref{eq:Zwanziger-gauge-eq1}) corresponds to the non-local gauge fixing in the conventional approach. 
In a perturbative sense, this is the reason why the self-interaction of the gauge field in the stochastic gauge fixing term simulates the contribution of the ghost in the conventional approach. The gauge fixing function (\ref{eq:Zwanziger-gauge-eq1}) ( and also (\ref{eq:FP-equilibrium-distribution-eq4})) provides a restoring force for the longitudinal mode of the gauge field, and therefore, it is expected in perturbative calculations that the probability distribution relaxes to the Faddeev-Popov distribution in equilibrium.

In a non-perturbative sense, however, the Faddeev-Popov distribution is not positive definite, because ${\rm det}(\partial\cdot \nabla)$ is not positive due to the presence of Gribov copies, and it is not justified that the probability distribution at finite stochastic time, $P[\t; A]$, uniquely relaxes to the Faddeev-Popov distribution $P_{\rm FP}[A]$ in equilibrium. The stochastic gauge fixing procedure enables us to introduce a class of covariant gauges which are non-holonomic in general. Therefore, if it happens in the stochastic-time evolution of the gauge field that the configuration of the gauge field is confined to the Gribov region, where the Faddeev-Popov operator $-\partial\cdot \nabla(A)$ is positive, the probability distribution $P(\t)$ uniquely relaxes to the equilibrium distribution $P_{\rm FP}$ for which $the$ $support$ $of$ $the$ $distribution$ $is$ $restricted$ $to$ $the$ $Gribov$ $region$.  Unfortunately, it is unclear whether this is indeed the case or not\cite{HHS}. In Ref.\citen{Zwanziger2}, it is argued from a slightly different point of view that the solution of (\ref{eq:FP-equilibrium-distribution-eq2}) with (\ref{eq:Zwanziger-gauge-eq1}) in the limit of the Landau gauge ($\a \rightarrow +0$) can be defined by the Faddeev-Popov distribution in the Landau gauge for which the support is restricted to the Gribov region 
$
\Omega \equiv \{ 
A^a_\mu (x) | \partial\cdot A =0,\ 
- \partial \cdot \nabla(A) > 0 
    \} .
$
In the Langevin equation (\ref{eq:YM4-Langevin-eq1}) with (\ref{eq:Zwanziger-gauge-eq1}), we may ignore the term $\delta S_{\rm YM}/ \delta A$ in the limit 
$\a \rightarrow + 0$. Then we obtain 
\bea
\Delta (\partial\cdot A)
= \Delta\t \displaystyle{\frac{1}{\a}} \partial\cdot \nabla (A) (\partial\cdot A) + \partial\cdot \Delta\eta        \ . 
\ena
Therefore, if the configuration of the gauge field initially exists inside the Gribov region $\Omega$, the random walk along the gauge orbit is strongly suppressed in the singular limit of the restoring force ($\a \rightarrow +0$),  while there exists at least one unstable direction along the gauge orbit outside of $\Omega$.\cite{Zwanziger2} We note that the restoring force is non-singular along the transverse direction. In this regard, on the basis of the Langevin equation, it is unclear whether the random walk of the gauge field configuration along the transverse direction, i.e. $\{A + \Delta A| \partial\cdot A = \partial\cdot \Delta A = 0 \}$, is restored enough to be confined to $\Omega$. In addition,  
since there are Gribov copies inside the Gribov region,\cite{STSF} in a precise resolution of the Gribov problem, the integration must be restricted to the so-called fundamental modular region, which is a proper subset of the Gribov region and free of Gribov copies.\footnote{
It is also conjectured in Ref.\citen{Zwanziger2} that the Landau gauge Faddeev-Popov distribution restricted to $\Omega$ gives the same correlation functions, 
$
\langle A(x_1)A(x_2)...A(x_n)
\rangle          \   , 
$ 
as those defined by the Landau gauge Faddeev-Popov distribution restricted to the fundamental modular region. 
} 
Although there have been some proposals for defining a globally valid path-integral for YM$_4$,\cite{AM,HK} the relations between these proposals and the stochastic gauge fixing procedure are also open questions. 

In the case of SYM$_4$, it is plausible that we encounter the Gribov problem provided that we work with a class of covariant gauges in the Wess-Zumino gauge. Although this may cause a problem similar to that which occurs in YM$_4$, it is beyond the scope of this paper to study the Gribov problem in SYM$_4$. In this paper, we do not investigate the Gribov problem in terms of our choice of the stochastic gauge fixing for SYM$_4$ and restrict our argument to showing the formal equivalence of the stochastic gauge fixing procedure and the conventional $naive$ Faddeev-Popov prescription by specifying the stochastic gauge fixing function in such a way that the Faddeev-Popov distribution of SYM$_4$ satisfies the stationary condition of the probability distribution in equilibrium. The main purpose of \S 3 and \S 4 is to define the stochastic gauge fixing procedure for SYM$_4$ and to derive the supersymmetric extension of Eqs.(\ref{eq:YM4-Langevin-eq1}), (\ref{eq:Zwanziger-gauge-eq1}), (\ref{eq:FP-equilibrium-distribution-eq2}) and (\ref{eq:FP-equilibrium-distribution-eq3}).

In order to define the stochastic gauge fixing procedure for SYM$_4$, we first introduce the stochastic-time $dependent$ auxiliary superfields $\Lambda$ and $\Lambda^\dagger$, which are chiral and anti-chiral, respectively, into the Langevin equation by performing the (inverse) local gauge transformation 
$e^{2gV}
\rightarrow 
e^{ig \Lambda^\dagger} e^{2gV} e^{-ig \Lambda}  $.  
Then we redefine the auxiliary chiral and anti-chiral superfields as   
\bea
\label{eq:SYM-auxiliaryfield-eq1}
\phi 
\equiv 
\displaystyle{\frac{1}{\beta}}
\displaystyle{\frac{i}{g}} \displaystyle{\frac{e^{ - ig L_{\Lambda}} - 1}{L_{\Lambda}}} 
{\dot  \Lambda } \ , \
{\bar \phi} 
\equiv 
\displaystyle{\frac{1}{\beta}}
\displaystyle{\frac{i}{g}} \displaystyle{\frac{e^{ - ig L_{\Lambda^\dagger}} - 1}{L_{\Lambda^\dagger}}} 
{\dot \Lambda}^\dagger      \ . 
\ena
Here we have denoted 
$
{\dot \Lambda} 
\equiv \displaystyle{\frac{\Delta \Lambda}{\Delta \t}}
$. 
The redefined auxiliary superfields, $\phi$ and ${\bar \phi}$, are also chiral and anti-chiral, 
${\bar D}\phi = D{\bar \phi} = 0$. 
Equation (\ref{eq:SYM-Langevin-eq1}) becomes 
\bea
\label{eq:SYM-Langevin-eq3} 
& {} & 
\displaystyle{\frac{1}{2g}} 
( \Delta e^{2gV} ) e^{-2gV}  
+ \displaystyle{\frac{i}{2}} \beta \Delta \t ( {\bar \phi} - e^{2g L_V} \phi  )    \nn 
& {} & \qquad\qquad 
= - \beta \Delta \t
\displaystyle{\frac{1}{2 g}} ( 
e^{2g L_V} {\cal D}{\cal W} + {\overline{\cal D}}{\overline{\cal W}} ) 
+ \Delta w       \ . 
\ena 
This Langevin equation is invariant under the stochastic-time $dependent$ local gauge transformation defined by 
\bea
\label{eq:extended-local-gauge-tr-eq1}
e^{2gV ( \t, z )} 
& \rightarrow & e^{-i g \Sigma^\dagger ( \t, z )}e^{2gV ( \t, z )}e^{ig \Sigma ( \t, z )}   \ ,  \nn
e^{ig \Lambda ( \t, z ) }        
& \rightarrow &
e^{ig \Lambda ( \t, z ) }  e^{ig \Sigma ( \t, z ) }    \ .
\ena
Here, the stochastic-time dependent transformation parameters, $\Sigma$ and $\Sigma^\dagger$, are chiral and anti-chiral, respectively. 
 Under this extended local gauge transformation (\lq\lq extended\rq\rq means stochastic-time $dependent$), the Langevin equation is covariantly transformed. In particular, the auxiliary superfields are transformed as 
\bea
\label{eq:extended-local-gauge-tr-eq2}
\phi 
& \rightarrow & 
\displaystyle{\frac{1}{\beta}}
 \displaystyle{\frac{i}{g}} \displaystyle{\frac{e^{- ig L_{\Sigma}} - 1}{L_{\Sigma}}} 
{\dot \Sigma}  
+ e^{- ig L_\Sigma} \phi   \ , \nn
{\bar \phi} 
& \rightarrow & 
\displaystyle{\frac{1}{\beta}}
 \displaystyle{\frac{i}{g}} \displaystyle{\frac{e^{ -ig L_{\Sigma_\dagger}} - 1}{L_{\Sigma^\dagger}}} 
{\dot \Sigma}^\dagger  
+ e^{-ig L_{\Sigma^\dagger}} {\bar \phi}   \ .   
\ena

The extended local gauge invariance of the Langevin equation (\ref{eq:SYM-Langevin-eq3}) under (\ref{eq:extended-local-gauge-tr-eq1}) and (\ref{eq:extended-local-gauge-tr-eq2}) must be broken by the gauge fixing procedure.   
In order to fix the gauge , we specify the gauge fixing functions, $\phi$ and ${\bar \phi}$, as follows: 
\bea
\label{eq:SYM-gauge-fixing-eq1}
\phi 
 =  i \displaystyle{\frac{\xi}{4}} 
 {\bar D}^2 D^2 V  , \ 
{\bar \phi}
 =  - i \displaystyle{\frac{\xi}{4}} 
 D^2 {\bar D}^2 V  \ . 
\ena
This choice of the gauge fixing functions is determined almost uniquely by dimensional analysis if we require that the Langevin equation is manifestly supersymmetric, invariant under the 4-dimensional Lorentz transformation and that the gauge fixing functions are linear with respect to the vector superfield $V$. 
In general, we may choose 
\bea
\label{eq:SYM-gauge-fixing-eq2}
\phi 
 =  i \displaystyle{\frac{\xi}{4}} 
 {\bar D}^2 D^2 h(V)  , \ 
{\bar \phi}
 =  - i \displaystyle{\frac{\xi}{4}} 
 D^2 {\bar D}^2 h(V)  \ ,
\ena
with an arbitrary function $h$ to specify the $local$ gauge fixing term. 
The Langevin equation (\ref{eq:SYM-Langevin-eq3}) with the gauge fixing functions (\ref{eq:SYM-gauge-fixing-eq1}) is a supersymmetric extension of the gauge fixed version of the YM Langevin equation. In the component expansion,\cite{Nakazawa1} (\ref{eq:SYM-Langevin-eq3}) with (\ref{eq:SYM-gauge-fixing-eq1}) includes the expression of the YM Langevin equation with the Zwanziger gauge fixing term (\ref{eq:Zwanziger-gauge-eq1}). Therefore, it provides the stochastic gauge fixing that is the supersymmetric extension of the so-called Zwanziger gauge in the YM case. We show in the next section that the $local$ gauge fixing functions (\ref{eq:SYM-gauge-fixing-eq1}) are the leading terms of the $non$-$local$ stochastic gauge fixing functions which are equivalent to the manifestly Lorentz covariant $local$ gauge fixing in the conventional Faddeev-Popov prescription. The perturbative aspect of this gauge fixing procedure is studied in \S 5.  For example, the superpropagator of the vector superfield, given in (\ref{eq:SYM-super-propagator}), is identical to that derived from the Langevin equation (\ref{eq:SYM-Langevin-eq3}) with (\ref{eq:SYM-gauge-fixing-eq1}) in the Feynman gauge $\xi = 1$, excluding the initial condition dependence of the superpropagator in solving the Langevin equation. 

Next, we show that the gauge fixing procedure we have introduced does not affect the time development of the local gauge invariant observables. This can be shown as follows. Let us consider the time development of a functional of $e^{2gV}$, $F[ e^{2gV} ]$. The time evolution equation is given by 
\bea
\label{eq:SYM-time-development-eq1} 
& {} &\!\!\!\!\!\! 
\Delta F[ e^{2gV} ]     \nn
& = &
 \beta \Delta \t 4g^2 
 \int\!\!\! d^8z \Big(  
  {\hat {\cal E}}^{(L)} (z)  
-  {\hat {\cal E}}^{(L)}(z) S  
- \displaystyle{\frac{i}{4g}} ( {\bar \phi} - e^{2g L_V} \phi )   
\Big)^a {\hat {\cal E}}^{(L)}_a (z) F[ e^{2gV} ]          .  
\ena 
In this expression, the contribution of the auxiliary fields defines the infinitesimal local gauge transformation. The generator of the infinitesimal local gauge transformation is defined by 
\bea
\label{eq:SYM-Gauss-law-eq1}
G ( \phi, {\bar \phi} ) 
& \equiv & 
 i \int\!\!\! d^8z  \Big(    
 {\bar \phi} -  e^{2g L_V} \phi        
 \Big)^a {\hat {\cal E}}^{(L)}_a ( z )            \nn 
& = & 
 i \int\!\!\! d^8z  \Big(    
( {\bar \phi} )^a {\hat {\cal E}}^{(L)}_a ( z ) -  ( \phi )^a {\hat {\cal E}}^{(R)}_a ( z )   \Big)     
 \ . 
\ena
This satisfies the supersymmetric extension of the Gauss law algebra defined 
by 
\bea
\label{eq:SYM-Gauss-law-eq2}
\Big{[} G( \phi_1,\ {\bar \phi}_1 ),  G( \phi_2,\ {\bar \phi}_2 ) \Big{]} 
= - i G ( [ \phi_1 , \phi_2],\ [ {\bar \phi}_1 , {\bar \phi}_2]   )       \ . 
\ena
Therefore, the introduction of the stochastic gauge fixing term has no effect on the time development of local gauge invariant observables, while it breaks the extended local gauge invariance, (\ref{eq:extended-local-gauge-tr-eq1}) and (\ref{eq:extended-local-gauge-tr-eq2}). This is the reason why the self-interaction of the vector superfield in the stochastic gauge fixing term simulates the Faddeev-Popov ghost contribution in the path-integral method.  

Here we comment on the implication of the extended local gauge transformation, (\ref{eq:extended-local-gauge-tr-eq1}) and (\ref{eq:extended-local-gauge-tr-eq2}). In the weak coupling region, (\ref{eq:extended-local-gauge-tr-eq2}) reads 
$
\phi \rightarrow 
\phi + \beta^{-1} {\dot \Sigma} + ig [ \phi,\ \Sigma ]  \ . 
$
The variation indicates a covariant derivative of the gauge parameter $\Sigma$ with respect to the stochastic time. In this sense, we may interpret the extended local gauge transformation as a 5-dimensional local gauge transformation for which the stochastic time is identified with the 5-th dimensional coordinate and the chiral and the anti-chiral superfields play the role of the 5-th component of the gauge field. For the YM case, this 5-dimensional extension of the local gauge transformation in SQM was discussed in Refs.\citen{CH,KOT,Nakazawa2} and \citen{Baulieu-GZ}. We note that the extended local gauge symmetry in (\ref{eq:extended-local-gauge-tr-eq1}) and (\ref{eq:extended-local-gauge-tr-eq2}) is $not$ a 5-dimensional supersymmetric extension. The role of the stochastic time is distinct from that of the other 4-dimensional space-time coordinates, and the global supersymmetry is the 4-dimensional one.

\section{Equivalence to the Faddeev-Popov prescription}    

In the YM$_4$ case, as we explain in \S 3, the equivalence of the stochastic gauge fixing procedure and the Faddeev-Popov prescription in the path-integral method is shown on the basis of the BRST invariance of the Faddeev-Popov probability distribution (\ref{eq:FP-equilibrium-distribution-eq3}).\cite{BZ,HHS} 
Here we give a proof of the assertion that the stochastic gauge fixing procedure proposed in the previous section for SYM$_4$ in the superfield formalism reproduces the Faddeev-Popov distribution given in the path-integral method. 

To show the equivalence, we first define the Faddeev-Popov distribution of SYM$_4$ in the BRST invariant path-integral method.\cite{FP} Let us introduce the BRST operator 
\bea
\label{eq:SYM-BRST-operator-eq1}
{\hat \delta}_{\rm BRST} 
& = & 
\int \!\! d^8z \Big\{ 
 i {\bar B}^a \displaystyle{\frac{\delta}{\delta {{\bar c}_a}}} 
+ i B^a \displaystyle{\frac{\delta}{\delta c_a}} 
- \displaystyle{\frac{g}{2}} [c\times c]^a  \displaystyle{\frac{\delta}{\delta c^a}} 
- \displaystyle{\frac{g}{2}} [{\bar c}\times {\bar c}]^a  \displaystyle{\frac{\delta}{\delta {\bar c}^a}}       \nn 
& {} & \qquad + 
\Bigl( - \displaystyle{\frac{i}{2}} L^a_{\ b}c^b 
+ \displaystyle{\frac{i}{2}} {\bar c}^b L_b^{\ a} 
 \Bigr) \displaystyle{\frac{\delta}{\delta V^a}} 
   \Big\}             \ , 
\ena 
where 
$[ X\times Y ]^a \equiv f^{abc} X^b Y^c$. 
The Faddeev-Popov distribution is then given by 
\bea
\label{eq:SYM-FP-weight-eq1}
P_{\rm FP} 
& \equiv & \int_{\rm gh} e^{- S_{\rm tot} }    \ ,     \nn 
S_{\rm tot} 
& = & S + S_{\rm gf} + S_{\rm FP}                   \ , \nn
S_{\rm gf} + S_{\rm FP}
& = & \int\!\!\! d^8z {\hat \delta}_{\rm BRST} F      
+  {\rm Tr} \Big\{  
 \int\!\!\! d^6z Bf + \int\!\!\! d^6{\bar z} {\overline B}{\overline f}  + 2\xi \int\!\!\! d^8z {\overline f}f  
\Big\}   \ , 
\ena
where 
$
\int_{\rm gh} \equiv 
\int{\cal D}B{\cal D}{\bar B}{\cal D}c{\cal D}{\bar c}{\cal D}c'{\cal D}{\bar c}'{\cal D}f{\cal D}{\bar f}
$. Here 
$
\varphi \equiv ( B, c, c', f)  
$
and 
$
{\bar \varphi} \equiv ( {\bar B}, {\bar c}, {\bar c}', {\bar f}) 
$ 
are chiral and anti-chiral superfields, i.e., they satisfy 
${\bar D}\varphi = D{\bar \varphi} = 0$. $f$ and ${\bar f}$ are variables for the gauge averaging. 
The function $F$ for the gauge fixing is defined by 
\bea
\label{eq:SYM-FP-weight-eq2}
F = -i \int \!\!\!d^8z 
( c' + {\bar c}' )^a V^a      \  . 
\ena
 The partition function is given by 
$Z = \int \!\!\! {\cal D}V P_{\rm FP}$ in the BRST invariant path-integral formulation.

We show that the Faddeev-Popov distribution, $P_{\rm FP}$ given by (\ref{eq:SYM-FP-weight-eq1}), is a stationary solution of the gauge fixed Fokker-Planck equation
\bea
\label{eq:SYM-Fokker-Planck-eq3} 
{\dot P}[ \t, e^{2gV} ] 
& = & 
 4\beta g^2 \int\!\!\! d^8z {\hat {\cal E}}^{(L)}_a (z) \Big(  
 {\hat {\cal E}}^{(L)} (z)
+ ( {\hat {\cal E}}^{(L)}(z) S )      \nn 
& {} & \qquad\qquad\qquad 
+ \displaystyle{\frac{i}{4g}}  ( {\bar \phi} - e^{2g L_V} \phi )   
\Big)^a  P[ \t, e^{2gV} ]         \ ,  
\ena 
which is derived from the stochastic-time evolution equation of the expectation value of observables in (\ref{eq:SYM-time-development-eq1}) under the appropriate choice of the gauge fixing functions $\phi$ and ${\bar \phi}$. 
 Namely, the r.h.s. of the Fokker-Planck equation (\ref{eq:SYM-Fokker-Planck-eq3}) vanishes for $P_{\rm FP}$, provided that the gauge fixing functions are defined by  
\bea
\label{eq:stochastic-gauge-fixing-function}
\phi^a (z)
& \equiv & 
4 g^2 \int_{\rm gh} \int \!\!\!d^8z' c^a (z) \Big\{ 
{\hat {\cal E}}_b^{(R)} (z') ( {\hat {\cal E}}_b^{(R)} (z') F ) 
e^{- S_{\rm tot} } 
\Big\} P_{\rm FP}^{-1}   
                                      \ ,    \nn
{\bar \phi}^a (z)
& \equiv & 
4 g^2 \int_{\rm gh} \int \!\!\!d^8z' {\bar c}^a (z) \Big\{ 
{\hat {\cal E}}_b^{(L)} (z') ( {\hat {\cal E}}_b^{(L)} (z') F ) 
e^{- S_{\rm tot} } 
\Big\} P_{\rm FP}^{-1} 
   \    . 
\ena

The proof is as follows. 
Substituting the distribution $P_{\rm FP}$ into the Fokker-Planck equation (\ref{eq:SYM-Fokker-Planck-eq3}), the r.h.s. reads
\bea
\label{eq:Fokker-Planck-evaluation-eq1}
\int \!\!\! d^8z {\hat {\cal E}}_a^{(L)} (z) \Big\{ 
& - & 
\Big( 
{\hat {\cal E}}_a^{(L)} (z) \int_{\rm gh} ( {\hat \delta}_{\rm BRST} F ) 
\Big) e^{- S_{\rm tot} }    \nn 
& {} & \qquad\qquad\qquad 
+ \displaystyle{\frac{i}{4g}} ( {\bar \phi} (z) - e^{2gL_V}\phi (z) )^a 
P_{\rm FP}
\Big\}    \ .    
\ena
%
%
The first term in (\ref{eq:Fokker-Planck-evaluation-eq1}) reads
\bea
\label{eq:Fokker-Planck-evaluation-eq2}
& {} & 
-\displaystyle{\frac{1}{2}}\int_{\rm gh} \int \!\!\! d^8z {\hat {\cal E}}_a^{(L)} (z) 
\Big\{  
 g f^{abc} {\bar c}^b ({\hat {\cal E}}_c^{(L)} (z) F ) + {\hat \delta}_{\rm BRST} ({\hat {\cal E}}_a^{(L)} (z) F) \Big\} 
 e^{- S_{\rm tot} }       \nn  
& {} & \ 
-\displaystyle{\frac{1}{2}}\int_{\rm gh} \int \!\!\! d^8z {\hat {\cal E}}_a^{(R)} (z) 
\Big\{  
 g f^{abc} c^b ({\hat {\cal E}}_c^{(R)} (z) F ) + {\hat \delta}_{\rm BRST} ({\hat {\cal E}}_a^{(R)} (z) F) \Big\} 
 e^{- S_{\rm tot} }      \ , 
\ena
%
%
while the second term in (\ref{eq:Fokker-Planck-evaluation-eq1}) is evaluated as 
\bea
\label{eq:Fokker-Planck-evaluation-eq3}
& {} & 
ig \int_{\rm gh} \int \!\!\!d^8z d^8z' {\bar c}^a (z) {\hat {\cal E}}_a^{(L)} (z)  \Big\{ 
{\hat {\cal E}}_b^{(L)} (z') ( {\hat {\cal E}}_b^{(L)} (z') F ) 
e^{- S_{\rm tot} }   
\Big\}         \nn 
 & {} & \qquad\quad 
 - ig  
\int_{\rm gh} \int \!\!\! d^8z d^8z'  c^a (z) {\hat {\cal E}}_a^{(R)} (z) \Big\{ 
{\hat {\cal E}}_b^{(R)} (z') ( {\hat {\cal E}}_b^{(R)} (z') F )   
e^{- S_{\rm tot}  }   
\Big\}             \nn
& = & 
\displaystyle{\frac{1}{2}} 
\int_{\rm gh} \int \!\!\! d^8z {\hat \delta}_{\rm BRST}  \Big\{ 
{\hat {\cal E}}_a^{(L)} (z) ( {\hat {\cal E}}_a^{(L)} (z) F ) 
e^{- S_{\rm tot} }   \nn
& {} & \qquad\qquad\qquad\qquad\qquad 
+
{\hat {\cal E}}_a^{(R)} (z) ( {\hat {\cal E}}_a^{(R)} (z) F )   
e^{- S_{\rm tot} }   
\Big\}             \nn
& = & 
 \displaystyle{\frac{1}{2}} 
\int_{\rm gh} \int \!\!\! d^8z 
\Big\{  
\Big[  {\hat \delta}_{\rm BRST},  \ 
{\hat {\cal E}}_a^{(L)} (z) \Big]  ( {\hat {\cal E}}_a^{(L)} (z) F ) 
e^{- S_{\rm tot} }     \nn
& {} & \qquad\qquad\qquad\qquad\qquad 
+ 
\Big[  {\hat \delta}_{\rm BRST},   \  
{\hat {\cal E}}_a^{(R)} (z) \Big] ( {\hat {\cal E}}_a^{(R)} (z) F )   
e^{- S_{\rm tot} }   
\Big\}                     \nn 
& {} & \qquad 
+  \displaystyle{\frac{1}{2}} 
\int_{\rm gh} \int \!\!\! d^8z 
\Big\{  
{\hat {\cal E}}_a^{(L)} (z)  {\hat \delta}_{\rm BRST} 
( {\hat {\cal E}}_a^{(L)} (z) F ) 
e^{- S_{\rm tot} }   \nn 
& {} & \qquad\qquad\qquad\qquad\qquad\qquad
+ 
{\hat {\cal E}}_a^{(R)} (z)  {\hat \delta}_{\rm BRST} 
( {\hat {\cal E}}_a^{(R)} (z) F )   
e^{- S_{\rm tot} }   
\Big\}              \nn
& = & - \ {\rm the\ first\ term\ in\ (\ref{eq:Fokker-Planck-evaluation-eq1})}            \ .      \qquad   {\rm Q.E.D.} 
\ena
In the above proof, we have used the commutation relations
\bea
[ {\hat \delta}_{\rm BRST},\ {\hat {\cal E}}^{(L)}_a (z) ]  
& = & 
-g f^{abc} {\bar c}^b (z){\hat {\cal E}}^{(L)}_c (z) \ , \nn
{[} {\hat \delta}_{\rm BRST},\ {\hat {\cal E}}^{(R)}_a (z) ]  
& = & 
-g f^{abc} c^b (z) {\hat {\cal E}}^{(R)}_c (z) \ .  \
\ena

Hence, the distribution $P_{\rm FP}$ is a stationary solution of the Fokker-Planck equation. 
However, as we have already mentioned, this does not imply the existence of the unique 
 equilibrium limit: 
$
\lim_{\t \rightarrow \infty }P ( \t ) = P_{\rm FP}
$. 
To demonstrate the relaxation of the probability distribution to a unique equilibrium distribution, we need a regularization procedure in order to reduce the number of the degrees of freedom that preserves the global supersymmetry as well as the local gauge symmetry. Furthermore, if the Gribov problem exists in the covariant gauge fixing (\ref{eq:SYM-FP-weight-eq2}) in the conventional approach, the positivity of $P_{\rm FP}$ is broken. Then, it is not justified that the probability distribution at finite stochastic time $P(\t)$ uniquely relaxes to the Faddeev-Popov distribution $P_{\rm FP}$ in equilibrium. We leave the investigation of these delicate problems to future works. 

The gauge fixing functions (\ref{eq:stochastic-gauge-fixing-function}), which are essential for the  proof of the equivalence, are different from those in (\ref{eq:SYM-gauge-fixing-eq1}), which are convenient for actual calculations. The proof of the equivalence relies on the particular form of the gauge fixing functions given in (\ref{eq:stochastic-gauge-fixing-function}). Before ending this section, we comment on the relation between these two gauge fixings, (\ref{eq:stochastic-gauge-fixing-function}) and (\ref{eq:SYM-gauge-fixing-eq1}). The gauge fixing functions (\ref{eq:stochastic-gauge-fixing-function}) are expressed as 
\bea
\label{eq:evaluation-gauge-fixing-function-eq1} 
\phi^a 
& = & 
2ig \int\!\!\! d^8z' \langle c^a (z) (c' + {\bar c}')^b (z') \rangle
     L^c_{\ b} (z') {\hat {\cal E}}^{(R)}_c (z') (S + S'_{\rm gf})  \nn
& {} & \qquad 
- 2ig \int\!\!\! d^8z'  L^c_{\ b} (z') {\hat {\cal E}}^{(R)}_c (z') \langle c^a (z) (c' + {\bar c}')^b (z')  \rangle          \  ,    \nn
{\bar \phi}^a 
& = & 
2ig \int\!\!\! d^8z' \langle {\bar c}^a (z) (c' + {\bar c}')^b (z') \rangle
     L_c^{\ b} (z') {\hat {\cal E}}^{(L)}_c (z') (S + S'_{\rm gf})  \nn
& {} & \qquad 
- 2ig \int\!\!\! d^8z'  L_c^{\ b} (z') {\hat {\cal E}}^{(L)}_c (z') \langle {\bar c}^a (z) (c' + {\bar c}')^b (z')  \rangle          \  , 
\ena
where 
$
S'_{\rm gf} \equiv \int\!\! d^8z \displaystyle{\frac{\xi}{8}} (D^2 V)^a({\bar D}^2 V)^a 
$, which is obtained by the integration of the auxiliary superfields $B$, ${\bar B}$, $f$ and ${\bar f}$ in the definition of $P_{\rm FP}$ (\ref{eq:SYM-FP-weight-eq1}). 
The ghost superpropagators in (\ref{eq:evaluation-gauge-fixing-function-eq1}) are defined by $P_{\rm FP}$, for example, 
\bea
\label{eq:evaluation-gauge-fixing-function-eq2} 
\langle {\bar c}_a (z){c'}^b (z') \rangle 
\equiv P_{\rm FP}^{-1} \int_{\rm gh} {\bar c}_a (z){c'}^b (z') e^{- S_{\rm tot}}      \  .
\ena
 We can expand each quantity in (\ref{eq:evaluation-gauge-fixing-function-eq1}) with respect to the coupling constant $g$ ( in other words, with respect to the vector superfield $V$ ). In principle, it is possible to evaluate (\ref{eq:evaluation-gauge-fixing-function-eq1}) perturbatively with the appropriate renormalization procedure of the singularities. In order to determine the leading term, we substitute the free propagators of the Faddeev-Popov ghost, 
\bea
\label{eq:evaluation-gauge-fixing-function-eq3} 
\langle {\bar c}'_a (z) c^b (z') \rangle^{(0)} 
& = & 
\langle {\bar c}_a (z) {c'}^b (z') \rangle^{(0)}    \ , \nn
& = & 
2 (-\displaystyle{\frac{1}{4}} D^2)_z (-\displaystyle{\frac{1}{4}} {\bar D}^2)_{z'} \langle a, z | (- \square)^{-1} | z', b \rangle       \ ,
\ena
into (\ref{eq:evaluation-gauge-fixing-function-eq1}). Then, we obtain 
\bea
\phi 
 & = & i \displaystyle{\frac{\xi}{4}} 
 {\bar D}^2 D^2 V  + {\cal O}(V^2)    \ , \nn 
{\bar \phi}
 & = & - i \displaystyle{\frac{\xi}{4}} 
 D^2 {\bar D}^2 V   + {\cal O}(V^2)    \ . 
\
\ena
This is what we alluded to in the previous section. 
Namely, the $local$ stochastic gauge fixing functions (\ref{eq:SYM-gauge-fixing-eq1}), which are almost unique from the dimensional analysis, are the linear terms in the stochastic gauge fixing functions that are equivalent to the conventional Faddeev-Popov prescription. This fact also implies that the stochastic gauge fixing term with (\ref{eq:SYM-gauge-fixing-eq1}) introduces a drift force for the longitudinal mode of the gauge field in the superfield Langevin equation (\ref{eq:SYM-Langevin-eq3}). In general, as is clear from (\ref{eq:evaluation-gauge-fixing-function-eq1}), the $local$ gauge fixing in the conventional approach yields $non$-$local$ stochastic gauge fixing functions. On the other hand, if we chose the stochastic gauge fixing functions to be $local$, the corresponding gauge fixing would be $non$-$local$ in the conventional approach. 
As we have shown in (\ref{eq:SYM-time-development-eq1}), the expectation values and the correlations for local gauge invariant observables are independent of the gauge fixing functions, and the difference between these two gauge fixings is irrelevant $in$ $a$ $perturbative$ $sense$.   

\section{Stochastic action for SYM$_4$ and the BRST symmetry} 

For perturbative analysis in SQM including the renormalization procedure, it is convenient to introduce the so-called stochastic action by the path-integral representation of the SQM approach.\cite{Gozzi,NY,CH,Okano} The construction of the stochastic action is standard. For SYM$_4$, applying a general formulation for the first-class constraint systems,\cite{Nakazawa2} the BRST invariant stochastic action is defined as follows. 

We first consider the continuum limit of the Langevin equation (\ref{eq:SYM-Langevin-eq3}) and the noise correlation (\ref{eq:SYM-Langevin-eq1}) by taking $\Delta \t \rightarrow 0$. This limiting procedure defines the continuum Langevin equation $E(V(\t, z)) = \eta(\t,z)$, where the noise superfield is defined by $\eta (\t, z) \equiv \lim_{\Delta\t \rightarrow 0} \Delta w (\t, z)/\Delta\t $, and its correlation is given by the continuum limit of (\ref{eq:SYM-Langevin-eq1}).  $E(V)$ is defined by the continuum limit of (\ref{eq:SYM-Langevin-eq3}). Then we insert unity in the form 
\bea
\label{eq:unity-eq1}
1 = \int\!\! \sqrt{G}{\cal D}V {\cal D}\phi {\cal D}{\bar \phi}  
\delta( E(V) - \eta ) {\rm det}\Big( 
\displaystyle{\frac{\delta E}{\delta V}}  
\Big) 
\delta(\Phi)\delta({\bar \Phi})  
\ena 
into the Gaussian integral representation of the correlation of the noise superfield $\eta$. Here, the functions $\Phi$ and ${\bar \Phi}$ are defined by 
\bea
\label{eq:unity-gauge-fixing-eq1}
\Phi \equiv \phi - {i\over4}\xi{\bar D}^2D^2 V  \ , 
\quad 
{\bar \Phi} \equiv {\bar \phi} + {i\over4}\xi D^2{\bar D}^2 V    \ , 
\ena
for the stochastic gauge fixing (\ref{eq:SYM-gauge-fixing-eq1}). 
In the It${\bar {\rm o}}$ stochastic calculus, we choose $\theta(0) = 0$, so that the dynamical variables are not correlated with the equal stochastic-time noise variables for the solution of the Langevin equation. This gives a trivial Jacobian, 
$ {\rm det}({\delta E}/{\delta V}) = 1$, 
while a careful evaluation of the integration measure\cite{NOSY} yields the contribution 
\bea
{\rm exp}\Bigl( {1\over 2}\int \!\!\! d^8z d\t G^{ab}(z,\t){\delta \over\delta V^a (z,\t)}{\delta S\over \delta V^b (z,\t)} \Bigr)       \  , 
\ena 
which includes $\delta^8 (0)$ in this case. Employing the regularization by dimensional reduction,\cite{Siegel,CJN}\ we discard this contribution. 

The integral representation of the $\delta$-functional in (\ref{eq:unity-eq1}) requires an auxiliary superfield ${\tilde \pi}^a$ that is not a vector superfield. By integrating out the noise superfield $\eta$, we obtain a stochastic action as a functional of the vector superfield $V^a$, the auxiliary superfield ${\tilde \pi}^a$, and the chiral and anti-chiral superfields $\phi$ and ${\bar \phi}$. After the redefinition of the auxiliary field ${\tilde \pi}^a \equiv L^a_{\ b} \varpi^b$, $\varpi^a$ becomes a vector superfield, ${\varpi^a}^\dagger = \varpi^a$.\footnote{
The hermiticity of $\varpi_a$ is defined only from the consistency of the transformation property. In analogy to the Ostarwalder-Schrader conjugation in the Euclidean space-time, the precise definition of the hermitian conjugation includes the time reversal with respect to the stochastic time,\cite{KM} 
$
V(\t)^{a\dagger} = V(-\t)^a , \
\varpi(\t)_a^\dagger = 
\varpi_a (-\t) - i \delta S/\delta V^a ( - \t) , 
$
$for$ $the$ $hermiticity$ $of$ $the$ $stochastic$ $hamiltonian$, 
$H \equiv i\int\!\!\! d^8z \t \varpi_a {\dot V}^a - K$. 
} 

As we have shown in (\ref{eq:extended-local-gauge-tr-eq1}), this stochastic action is also invariant under the extended local gauge transformation, provided that the auxiliary superfield ${\tilde \pi}$ is transformed as 
${\tilde \pi}^a \rightarrow e^{-i g L_{\Sigma^\dagger}} {\tilde \pi}^a$.   
This also induces the local gauge transformation of the auxiliary vector superfield $\varpi$. 
Therefore, we can introduce a BRST transformation by making the replacement 
$ 
\Sigma \rightarrow - \lambda c, \ 
\Sigma^\dagger \rightarrow - \lambda {\bar c}. 
$ 
Here, $\lambda$ is a purely imaginary Grassmann number. 
The BRST symmetry is realized on the field contents  
$\{
 ( V^a, \varpi_a ),\ ( \phi^a, {\tilde B}^a, c^a, {\tilde c}_a ),\ 
 ( {\bar \phi}^a, {\bar {\tilde B}}^a, {\bar c}^a, {\bar {\tilde c}}_a ) 
 \}$. 
 Here, 
$\phi^a$, ${\tilde B}^a$, $c^a$ and ${\tilde c}^a$ are chiral and 
 ${\bar \phi}^a$, ${\bar {\tilde B}}^a$, ${\bar c}^a$ and ${\bar {\tilde c}}_a$ are anti-chiral 
 superfield. As is clear from (\ref{eq:unity-eq1}), we impose the gauge conditions (\ref{eq:SYM-gauge-fixing-eq1}) with the Nakanishi-Lautrup fields, ${\tilde B}$ and ${\bar {\tilde B}}$, and the functions, $\Phi$ and ${\bar \Phi}$.  
This choice of the gauge conditions is an extension of the \lq\lq flow gauges \rq\rq in the YM case, in which the Faddeev-Popov ghost decouples from the gauge field.\cite{CH,KOT}  
Explicitly, the extended (or 5-dimensional) BRST transformation is defined by 
\bea
\label{eq:stochastic-BRST-tr.-eq1}
\delta_{\rm BRST} V^a 
& = & 
- \lambda \displaystyle{\frac{i}{2}} L^a_{\ b}c^b 
+ \lambda \displaystyle{\frac{i}{2}} {\bar c}^b L_b^{\ a}   \ , \nn 
\delta_{\rm BRST} \varpi_a 
& = & 
 \lambda \displaystyle{\frac{i}{2}} \pa_a L^c_{\ b} \varpi_c c^b 
- \lambda \displaystyle{\frac{i}{2}} {\bar c}^b \pa_a L_b^{\ c} \varpi_c  
      \ , \nn 
\delta_{\rm BRST} \phi^a 
& = &   \lambda 2\kappa {\dot c}^a  
+ \lambda g [\phi\times c]^a    \ , \nn 
\delta_{\rm BRST} {\bar \phi}^a 
& = &    \lambda 2\kappa {\dot {\bar c}}^a  
+ \lambda g [{\bar \phi}\times {\bar c}]^a       \ , \nn 
\delta_{\rm BRST} c^a 
& = & 
- \lambda \displaystyle{\frac{g}{2}} [c\times c]^a           \ , \nn
\delta_{\rm BRST} {\bar c}^a   
& = & 
- \lambda  \displaystyle{\frac{g}{2}} [{\bar c}\times {\bar c}]^a \ , \nn
\delta_{\rm BRST} {\tilde c}_a   
& = &  i \lambda {\tilde B}_a     \ , \qquad\quad 
\delta_{\rm BRST} {\bar {\tilde c}}_a
 =    i \lambda {\bar {\tilde B}}_a \ , \nn
\delta_{\rm BRST} {\tilde B}^a 
& = & 0   \ , \qquad\qquad\quad 
\delta_{\rm BRST} {\bar {\tilde B}}^a           
 =  0    \  . 
\ena
Here, we have redefined the scaling parameter of the stochastic time as $\beta \equiv - \displaystyle{\frac{1}{2\kappa}}$. (Also see Appendix A for the convention.)
The extended (or 5-dimensional) BRST invariant stochastic action is given by 
\bea
\label{eq:BRST-invariant-stochastic-action-eq1}
Z 
& = &  
\int d\mu e^{ K_{\rm BRST}}        \ , 
\quad 
d\mu 
= {\cal D}V{\cal D}\varpi{\cal D}c{\cal D}{\tilde {c}}{\cal D}{\bar c}{\cal D}{\bar {\tilde c}} {\cal D}\phi {\cal D}{\bar \phi} {\cal D}{\tilde B}{\cal D}{\bar {\tilde B}}  
\ , \nn 
K 
& \equiv & 
\!\!\! \int \!\!\! d^8z d\t \Big[ 
\displaystyle{\frac{1}{2\kappa}} G^{ab} \varpi_a \varpi_b 
 + i \varpi_a \Big\{  
{\dot V}^a - \displaystyle{\frac{1}{4g\kappa}} 
\Big(   
L^a_{\ b} ( {\cal D}^\a W_\a )^b  
+ ( {\bar {\cal D}}_{\dot \a} {\bar W}^{\dot \a} )^b L_b^{\ a} 
\Big)   
\Big\}                  \ , \nn 
& {} & \qquad\qquad
- \displaystyle{\frac{1}{4\kappa}} \varpi_a 
 ( L^a_{\ b} \phi^b - {\bar \phi}^b L_b^{\ a} )  \Big]       \ , \nn
K_{\rm BRST} 
& \equiv & 
K  + K_{\rm NL} + K_{\rm FP}      \  , \nn
K_{\rm NL} 
&+& 
K_{\rm FP}  
 \equiv  
- \displaystyle{\frac{1}{2\kappa}} \int \!\!\! d^8z d\t \Big\{ s 
\Big( 
{\tilde c}_a {\bar \Phi}^a 
+ {\bar {\tilde c}}_a \Phi^a   
 \Big)      \Big\}   .
\ena
Here, $\delta_{\rm BRST} {\cal O} \equiv \lambda (s{\cal O})$. $\Phi$ and ${\bar \Phi}$ are defined by (\ref{eq:unity-gauge-fixing-eq1}). This formulation is sufficient to perform the perturbative analysis. As we explain in the following, we may consider truncated versions of (\ref{eq:stochastic-BRST-tr.-eq1}) and (\ref{eq:BRST-invariant-stochastic-action-eq1}). 

We redefine the Nakanishi-Lautrup fields, $B$ and ${\bar B}$, and the anti-FP ghosts, ${\tilde c}$ and ${\bar {\tilde c}}$, as 
\bea
\label{eq:redefinition-FP-ghosts-eq1}
B 
& \equiv &
\Bigl( -\displaystyle{\frac{1}{4}} {\bar D}^2 \Bigr) {\bar {\tilde B}} \ ,  \quad 
{\bar B} 
\equiv 
\Bigl( -\displaystyle{\frac{1}{4}} D^2 \Bigr) {\tilde B}   \ , \nn
c' 
& \equiv &
\Bigl( -\displaystyle{\frac{1}{4}} {\bar D}^2 \Bigr) {\bar {\tilde c}} \ ,  \quad 
{\bar c}' 
\equiv 
\Bigl( -\displaystyle{\frac{1}{4}} D^2 \Bigr) {\tilde c}   \ . 
\ena
The BRST transformation for anti-FP ghosts becomes 
\bea
\label{eq:redefined-BTST-eq1}
\delta_{\rm BRST} c'_a   
 =   i \lambda B_a     \ , \qquad\quad 
\delta_{\rm BRST} {\bar c}'_a
 =    i \lambda {\bar B}_a \ . 
\ena
Correspondingly, the BRST exact part, $K_{\rm NL}$ and $K_{\rm FP}$ in (\ref{eq:BRST-invariant-stochastic-action-eq1}), is expressed in terms of the chiral measure in the superspace, 
$d^6z = d^4x d^2{\theta}$ and $d^6{\bar z} = d^4x d^2{\bar \theta}$, 
as 
\bea
\label{eq:redefined-BTST-eq2} 
K_{\rm NL} + K_{\rm FP} 
= 
- \displaystyle{\frac{1}{2\kappa}} \Big\{ s
\Big( 
\int \!\!\! d^6{\bar z} d\t {\bar c}'_a 
{\bar \Phi}^a 
+ 
\int \!\!\! d^6z d\t c'_a \Phi^a 
 \Big)   \Big\} . 
\ena

To reduce the number of auxiliary fields, we consider a consistent truncation of the BRST transformation (\ref{eq:stochastic-BRST-tr.-eq1}) by integrating out the auxiliary superfields, $\phi$ and ${\bar \phi}$, and the Nakanishi-Lautrup superfields, $B$ and ${\bar B}$. Since $\phi$ and ${\bar \phi}$ are chiral and anti-chiral, respectively, the integrations of these superfields yield $\delta$-functionals which impose the conditions
\bea
\label{eq:truncation-BRST-eq1}
 i {\bar B}_a 
& = & 
\Bigl( -\displaystyle{\frac{1}{4}} D^2 \Bigr)\Bigl( \displaystyle{\frac{1}{2}} \varpi_b L^{\ b}_a \Bigr) - g [{\bar c}\times {\bar c}']_a     \ , \nn
 i B_a 
& = & 
- \Bigl( -\displaystyle{\frac{1}{4}} {\bar D}^2 \Bigr) 
\Bigl( \displaystyle{\frac{1}{2}} \varpi_b L^b_{\ a} \Bigr) - g [c\times c']_a     \ .
\ena

The equations of motion (\ref{eq:truncation-BRST-eq1}) yield a truncated BRST symmetry, as well as a corresponding truncated BRST invariant stochastic action.   The truncated BRST transformation is expressed in the extended phase space $\Big\{ ( V^a, \varpi_a ), ( c^a, {c'}^a ), ( {\bar c}^a, {\bar c}^{'a} ) \Big\} $ as 
\bea
\label{eq:stochastic-BRST-tr.-eq2}
{\hat \delta}_{\rm BRST} V^a 
& = & 
- \lambda \displaystyle{\frac{i}{2}} L^a_{\ b}c^b 
+ \lambda \displaystyle{\frac{i}{2}} {\bar c}^b L_b^{\ a}   \ , \nn 
{\hat \delta}_{\rm BRST} \varpi_a 
& = & 
 \lambda \displaystyle{\frac{i}{2}} \pa_a L^c_{\ b} \varpi_c c^b 
- \lambda \displaystyle{\frac{i}{2}} {\bar c}^b \pa_a L_b^{\ c} \varpi_c  
 \ , \nn 
{\hat \delta}_{\rm BRST} c^a
& = & 
- \lambda \displaystyle{\frac{g}{2}} [c\times c]^a           \ , \nn
{\hat \delta}_{\rm BRST} {\bar c}^a   
& = & 
- \lambda  \displaystyle{\frac{g}{2}} [{\bar c}\times {\bar c}]^a \ , \nn
{\hat \delta}_{\rm BRST} {c'}_a   
& = & 
- \lambda \Bigl( -\displaystyle{\frac{1}{4}} {\bar D}^2 \Bigr) \Bigl( \displaystyle{\frac{1}{2}} \varpi_b L^b_{\ a}  \Bigr)
- \lambda g [c\times c']_a           \ , \nn 
{\hat \delta}_{\rm BRST} {{\bar c}'}_a
& = &  
 \lambda \Bigl( -\displaystyle{\frac{1}{4}} D^2 \Bigr) \Bigl( \displaystyle{\frac{1}{2}}  L_a^{\ b} \varpi _b \Bigr) 
 - \lambda g [{\bar c}\times {\bar c}']_a      
     \ .       
\ena
We note that the truncated BRST transformation is $nilpotent$. It preserves the chirality of the superfields; i.e., 
${\bar D} c = {\bar D} c' = D {\bar c} = D {\bar c}' = 0$. The truncated BRST invariant stochastic action is reduced from the 5-dimensional extended one (\ref{eq:BRST-invariant-stochastic-action-eq1}): 
\bea
\label{eq:BRST-invariant-stochastic-action-eq2}
Z 
& = &  
\int d{\hat \mu} e^{ {\hat K}_{\rm BRST}}        \ , \quad 
d\mu = {\cal D}V{\cal D}\varpi{\cal D}c{\cal D}c'{\cal D}{\bar c}{\cal D}{{\bar c}'}       \ ,  \nn 
{\hat K}_{\rm BRST} 
& \equiv & 
K' +  \int \!\!\! d^6z d\t  
 {c'}_a {\dot c}^a 
+  \int \!\!\! d^6{\bar z} d\t {{\bar c}'}_a {\dot {\bar c}}^a   \nn 
& {} & \qquad\qquad 
+  i \displaystyle{\frac{\xi}{2\kappa}}\int \!\!\! d^8z d\t \Big[ {\hat s}( {{\bar c}'}_a {\bar D}^2 V^a  -  {c'}_a D^2 V^a )      \Big]  \ .
\ena
Here, ${\hat \delta}_{\rm BRST} {\cal O} \equiv \lambda ({\hat s}{\cal O})$, and $K'$ is given by $K$ with $\phi={\bar \phi}=0$ in (\ref{eq:BRST-invariant-stochastic-action-eq1}). The BRST invariance of this truncated stochastic action is manifest, if we regard (\ref{eq:BRST-invariant-stochastic-action-eq2}) as a Legendre transformation, because of the invariance of the derivative terms with respect to the stochastic time. Namely, we have a non-trivial BRST cohomology of (\ref{eq:stochastic-BRST-tr.-eq2}); 
$
{\hat \delta}_{\rm BRST} \Big\{ 
i \int \!\! d^8z d\t \varpi_a {\dot V}^a  
+  
\int \!\! d^6z d\t {c'}_a {\dot c}^a 
+ 
 \int \!\! d^6{\bar z} d\t {{\bar c}'}_a {\dot {\bar c}}^a 
 \Big\} 
= 0 . 
$ 
We also note that the canonical conjugate momentum of the FP ghost $c$ (${\bar c}$) with respect to the stochastic time is $c'$ (${\bar c}'$). 
The BRST exact part in (\ref{eq:BRST-invariant-stochastic-action-eq2}) is decomposed as 
\bea
\label{eq:FP-ghost-sector}
& {} & 
i \displaystyle{\frac{\xi}{2\kappa}} \int \!\!\! d^8z d\t  \delta'_{\rm BRST} (  {{\bar c}'}_a {\bar D}^2 V^a  -  {c'}_a D^2 V^a )   \nn
& {} & \quad
= 
- i \displaystyle{\frac{\xi}{16\kappa}} \int \!\!\! d^8z d\t  
 \varpi_a  \Big( 
L_b^{\ a} D^2{\bar D}^2 V^b + L^a_{\ b} {\bar D}^2 D^2 V^b 
   \Big)     \nn 
& {} & \qquad 
+ i \displaystyle{\frac{\xi}{2\kappa}} \int \!\!\! d^8z d\t    
\Big\{ 
 \displaystyle{\frac{i}{2}} ({\bar c}'_a {\bar D}^2 - c'_a D^2 ) ( L_{\ b}^a c^b - {\bar c}^b L_b^{\ a} )    \nn
& {} & \qquad 
- g [{\bar c}\times {\bar c}']_a {\bar D}^2 V^a 
+ g [c\times c']_a D^2 V^a         \Big\}       \ . 
\ena
The first term on the r.h.s. of this expression is the stochastic gauge fixing term. The second and third terms describe the interactions between the vector superfield and the FP ghosts. We note that the truncated BRST symmetry (\ref{eq:stochastic-BRST-tr.-eq2}) is independent of the specific form of the stochastic gauge fixing functions. For example, it allows the gauge fixing functions given in (\ref{eq:SYM-gauge-fixing-eq2}). 

The stochastic action (\ref{eq:BRST-invariant-stochastic-action-eq2}) is also invariant under the scale transformation of Faddeev-Popov ghosts: 
\bea
\label{FPghost-scale-transf.}
c \rightarrow e^\rho c ,\ 
c' \rightarrow e^{-\rho} c' ,\ 
{\bar c} \rightarrow e^\rho {\bar c},\ 
{\bar c}' \rightarrow e^{-\rho} {\bar c}'\ . 
\ena
We assign the ghost number $+1$ for $c$ and ${\bar c}$, $-1$ for $c'$ and ${\bar c}'$, and $0$ for the remaining fields. 

In order to define the superpropagators, we determine the kinetic term by taking the limit $g \rightarrow 0$. It is given by
\bea
\label{eq:kinetic-term-eq1}
{\hat K}_{\rm BRST}^{(0)} 
& = & 
\!\!\! \int \!\!\! d^8z d\t \Big[ 
\displaystyle{\frac{1}{2\kappa}} \varpi_a \varpi^a 
 + i \varpi_a \Big\{  
{\dot V}^a + \displaystyle{\frac{1}{\kappa}} 
\Big(   
- \square + \displaystyle{\frac{1 - \xi}{16}} 
( {\bar D}^2 D^2 + D^2 {\bar D}^2 )
\Big) V^a  
\Big\}    
\Big]          \ , \nn 
& {} & 
+  \int \!\!\! d^6z d\t c'_a \Bigl( 
\displaystyle{\frac{\pa}{\pa \t}} - \displaystyle{\frac{\xi}{16\kappa}} {\bar D}^2 D^2 \Bigr) c^a    
+  \int \!\!\! d^6{\bar z} d\t {\bar c}'_a \Bigl( 
\displaystyle{\frac{\pa}{\pa \t}} - \displaystyle{\frac{\xi}{16\kappa}} D^2 {\bar D}^2 \Bigr) {\bar c}^a    \ .  
\ena
For example, in the Feynman gauge, $\xi = 1$, we obtain the simplest superpropagators: 
\bea
\label{eq:SYM-super-propagator}
\langle V^a ( \t, z ) V^b ( \t', z' ) \rangle 
& = & 
- \delta^{ab} \displaystyle{\frac{1}{\kappa}} 
\displaystyle{\frac{1}{( i\omega + k^2/\kappa )( - i\omega + k^2/\kappa )}}  
\delta^2 ( \theta - \theta' ) \delta^2 ( {\bar \theta} - {\bar \theta'} )         \ , \nn 
\langle V^a ( \t, z ) \varpi^b ( \t', z' ) \rangle 
& = & 
i \delta^{ab} 
\displaystyle{\frac{1}{(  i\omega + k^2/\kappa )}} 
\delta^2 ( \theta - \theta' ) \delta^2 ( {\bar \theta} - {\bar \theta'} )            \ , \nn 
\langle \varpi^a ( \t, z ) \varpi^b ( \t', z' ) \rangle 
& = & 0     \ ,             \nn
\langle c^a ( \t, z ) {c'}^b ( \t', z' ) \rangle      
& = & 
-  
\displaystyle{\frac{1}{(  i\omega + k^2/\kappa )}} 
\Bigl( - \displaystyle{\frac{1}{4}} {\tilde {\bar D}}^2 \Bigr)
\delta^2 ( \theta - \theta' ) \delta^2 ( {\bar \theta} - {\bar \theta'} )            \ , \nn
\langle {\bar c}^a ( \t, z ) {\bar c}^{'b} ( \t', z' ) \rangle
& = & 
-  
\displaystyle{\frac{1}{(  i\omega + k^2/\kappa )}} 
\Bigl( - \displaystyle{\frac{1}{4}} {\tilde D}^2 \Bigr)
\delta^2 ( \theta - \theta' ) \delta^2 ( {\bar \theta} - {\bar \theta'} )            \ .
\ena
Here, we have suppressed the momentum integration 
$\int \!\! d^4k d\omega (2\pi)^{-5}e^{ik \cdot ( x - x' ) + i \omega ( \t - \t' )}$. 
We have also defined the covariant derivatives 
${\tilde D}_\a \equiv \pa_\a - \sigma^m_{\a{\dot \a}} \theta^{\dot \a}k_m $ 
and 
${\tilde {\bar D}}_{\dot \a} \equiv - {\bar \pa}_{\dot \a} + \theta^\a \sigma^m_{\a{\dot \a}} k_m $. 
As is clear from these expressions in (\ref{eq:SYM-super-propagator}), the superpropagators for the vector superfields, $V$ and $\varpi$, are the same as those for the YM case, except the $\delta$-functions for the Grassmannian coordinates, while the superpropagators for F-P ghosts include the chiral projection operators, $D^2$ and ${\bar D}^2$.  We comment on the retarded (or causal) property of the superpropagators for $\langle V \varpi\rangle$ and the F-P ghosts. In the superfield Langevin equation, we have to choose $\theta(0)=0$, so that the dynamical fields are not correlated with the equal stochastic-time noise fields in terms of the It${\bar {\rm o}}$ stochastic calculus. This prescription is also a prerequisite in the definitions of the stochastic action and the superpropagators. Therefore, in the configuration space, the choice $\theta (0)=0$ is always assumed for the definition of $\langle V \varpi\rangle$ and the ghost superpropagators. 

The superpropagators and the BRST invariant stochastic action (\ref{eq:BRST-invariant-stochastic-action-eq2}) define the perturbative expansion with supergraphs in the SQM approach by preserving the global supersymmetry in the superfield formalism. We note that we have specified the regularization procedure by employing regularization by dimensional reduction to discard the singularities, such as $\delta^8 (0)$.

%
\section{Power counting and the renormalizability of SYM$_4$ \\ 
in the SQM approach} 

In the path-integral approach, the perturbative analysis is carried out in terms of the supergraphs. This ensures the non-renormalization theorem in SYM$_4$.\cite{FP,GSR} As we have explained, the perturbative calculation with the supergraphs can also be applied by using the superpropagators defined in (\ref{eq:SYM-super-propagator}) in the context of SQM. Therefore, the non-renormalization theorem also holds in the SQM approach, which is a consequence of the fact that the non-vanishing contributions in the perturbative expansion come from the loops which include the four spinor covariant derivatives $D_\a D_\b {\bar D}_{\dot \a} {\bar D}_{\dot \b}$ to eliminate $\delta^2(\theta-\theta')\delta^2({\bar \theta}- {\bar \theta}')|_{\theta=\theta', {\bar \theta}={\bar \theta}'}$ in the superspace $\delta$-function. Furthermore, the symmetries and the simple power counting argument indicate the multiplicative renormalizability of the SQM approach in terms of the BRST invariant stochastic action for SYM$_4$. 

Let us estimate the divergence of a Feynman diagram $G( V,I,E )$ with  $V$ vertices, $I$ internal lines, $E$ external lines and $L \equiv I - V +1$ loops. More precisely, we define, 
\bea
I \equiv I_{V \varpi} + I_{V V}, \qquad 
E &\equiv& E_\varpi + E_V.  \nn
E_\varpi : \#\ {\rm external\ lines\ of}\ \varpi, \quad 
E_V &:& \#\ {\rm external\ lines\ of}\ V             \nn
I_{V \varpi} : \#\ {\rm internal\ lines\ of}\ \langle V \varpi \rangle,  \quad 
I_{V V} &:& \#\ {\rm internal\ lines\ of}\ \langle V V \rangle     \nn 
V_{1,n}  :  \#\ {\rm vertices\ with}\ \varpi\ {\rm and}\ n\ V, \
V_{2,n} &:& \#\ {\rm vertices\ with\ two}\ \varpi\ {\rm and}\ n\ V . 
\ena
We also have the topological relations
\bea
\sum_{n=2} n ( V_{1,n} + V_{2,n} ) 
 =  
2 I_{V V} + I_{V \varpi} + E_V,      \quad
\sum_{n=2} ( V_{1,n} + 2 V_{2,n} ) 
 =  
I_{V \varpi} + E_\varpi           .  
\ena
The Feynman diagram $G( V,I,E )$ may be evaluated  as 
\bea
\label{eq:estimation-divergences-eq1}
& {} &\!\!\!\!\!\!  G( V, I, E )   \nn 
& \approx & \Bigl( \int \!\!\! d^4x_i d\t_i \Bigr) ^V \Bigl( \int \!\!\! d^2\theta_j d^2{\bar \theta}_j \Bigr) ^V  ( \langle V_i V_j \rangle )^{I_{VV}} ( \langle V_{i'} \varpi_{j'} \rangle )^{I_{V \varpi}} 
( V_{ex} )^{E_V} ( \varpi_{ex} )^{E_\varpi}               \nn 
& \approx & ( {\tilde V}_{ex} )^{E_V} ( {\tilde \varpi}_{ex} )^{E_{\varpi}} 
\delta^4 \Bigl( \sum p_{ex} \Bigr) \delta \Bigl( \sum \omega_{ex} \Bigr) 
\int \!\!\!d^2\theta d^2{\bar \theta}  
\Bigl( \int \!\!\! d^4k d\omega \Bigr) ^L 
\Bigl( \displaystyle{\frac{1}{k^2}} \Bigr) ^{I_{VV}} \nn
& {} &\!\!\!\!\!\! 
 \Bigl( \displaystyle{\frac{1}{\pm i \omega + k^2}} \Bigr) ^{I_{V V} + I_{V \varpi}}   \!\!
 \Bigl( \int d^2\theta d^2{\bar \theta} \Bigr) ^{V-1} \!\!\! ( \delta^2( \theta_{ij}) \delta^2( {\bar \theta}_{ij} ) )^I \Big( {\tilde D}_\a {\tilde D}_\b 
  {\tilde {\bar D}}_{\dot \c} {\tilde {\bar D}}_{\dot \delta} \Big)^{\sum_{n=2} V_{1,n}}       \!\!\!  .
\ena
Thus the degree of the ultra-violet divergence of the Feynman diagram $G(V,I,E)$ is estimated as  
\bea
\label{eq:estimation-divergences-eq2}
4L -2 I_{V V} -2 I + 2\Bigl( \sum_{n=2} V_{1,n} \Bigr)       
 =  4 - 2E_\varpi      \ .
\ena 

It is necessary to introduce two types of counter-terms to cancel the divergences, which are constrained by the Ward-Takahashi identities of the BRST symmetry (\ref{eq:stochastic-BRST-tr.-eq2}):\\
\noindent
(i) For $E_\varpi = 2$, the logarithmic divergences are cancelled by terms of the form $\varpi^2 V^n$, with $n=0,1,2,...\infty$.\\
\noindent
(ii) For $E_\varpi = 1$, the logarithmic divergences are cancelled by terms of the forms, such as $\varpi {\dot V}$ and $\varpi D^\a {\bar D}^2 D_\a V^n$, 
with $n=1,2,...\infty$. \\
\noindent
(iii) For $E_\varpi = 0$, there are no Feynman diagrams.

We must pay a special attention to the $E_\varpi = 0$ case. Since the superpropagator $\langle V \varpi \rangle$ is a retarded one with respect to the stochastic time, the contribution of a loop vanishes if it consists of only the superpropagator $\langle V \varpi \rangle$ as the internal lines. This means that the Feynman diagram includes at least one external line of $\varpi$. There is no ambiguity due to the regularization procedure with regard to this fact, because we have to choose $\theta (0)=0$ in the It${\bar {\rm o}}$ stochastic calculus. This is an essential point for the renormalizability of the present approach. Without this property, it is obvious that the SQM approach is not renormalizable. 
The same conclusion is obtained from the $stochastic$ Ward-Takahashi identity.\cite{ZZ} 
We note that, for $E_\varpi = 1$, the divergence appears to be quadratic. However, the BRST invariance reduces the order of the divergence to logarithmic. 

The power counting of the Feynman diagrams with FP ghost external lines is similar to what we have explained for the Feynman diagrams with only external vector superfields, except that the ghosts are the chiral and anti-chiral superfields. 
The degree of the divergence is given by 
\bea
\label{eq:estimation-divergences-eq3} 
4 - 2E_\varpi - \displaystyle{\frac{3}{2}} ( E_c + E_{c'} + E_{\bar c} + E_{{\bar c}'} )   \ , 
\ena
where $E_c$, $E_{c'}$, $E_{\bar c}$ and $E_{{\bar c}'}$ are the numbers of the external ghost lines for $c$, $c'$, ${\bar c}$ and ${\bar c}'$, respectively. 
We have also used the ghost number conservation to derive (\ref{eq:estimation-divergences-eq3}). 
Because of the retarded (or causal) nature of the FP ghost superpropagators in (\ref{eq:SYM-super-propagator}), the ghost loops do not contribute to the renormalization of the 1-P-I vertices. This means that, $in$ $the$ $actual$ $calculation$ $of$ $physical$ $quantities$ $such$ $as$ $the$ $\beta$-$function$ $of$ $the$ $gauge$ $coupling$, $we$ $may$ $discard$ $the$ $F$-$P$ $ghost$ $sector$. On the other hand, since the vector superfields couple with the ghosts and they contribute to the renormalization of FP ghost sector, (\ref{eq:FP-ghost-sector}) and 
$
 \int \!\! d\t ( 
\int \!\! d^6z {c'}_a {\dot c}^a 
+ \int \!\! d^6{\bar z} {{\bar c}'}_a {\dot {\bar c}}^a 
 ) 
$
are renormalized by the loop correction of the vector superfields. 
Therefore, the existence of the BRST symmetry is necessary for the formal proof of the manifest renormalizability of SYM$_4$ in this context, in order to apply the standard arguments in terms of the Ward-Takahashi identity. 
As we have already noted with regard to (\ref{eq:BRST-invariant-stochastic-action-eq2}), the non-trivial BRST cohomology 
$
{\hat \delta}_{\rm BRST} \{ \int \!\! d\t ( 
 i \int \!\! d^8z \varpi_a {\dot V}^a + 
\int \!\! d^6z {c'}_a {\dot c}^a 
+ \int \!\! d^6{\bar z} {{\bar c}'}_a {\dot {\bar c}}^a 
 ) \} = 0 
$ plays a key role for the renormalizability of the F-P ghost sector, and this is the reason why we have introduced the scaling parameter $\kappa$. The multiplicative renomalizability requires this additional coupling constant $\kappa$, due to the existence of this particular non-trivial BRST cohomology. Since the stochastic gauge fixing term and the remaining F-P ghost term takes a BRST exact form, the power counting argument, the BRST symmetry, and the ghost number conservation indicate that the stochastic action (\ref{eq:BRST-invariant-stochastic-action-eq2}) is multiplicatively renormalizable by the wave function renormalizations and the renormalizations of the couplings $g$, $\kappa$ and $\xi$. 

We add a comment concerning the proof given in Appendix B of the renormalizability of the SQM approach in terms of the BRST invariant stochastic action. In the truncated version (\ref{eq:BRST-invariant-stochastic-action-eq2}), since the truncated BRST symmetry is independent of the stochastic gauge fixing term, there is an ambiguity for the renormalization of the gauge fixing term, which cannot be determined by the symmetry argument. As we have already explained as an example, we can change the stochastic gauge fixing, replacing $V$ by $h(V)$ with an arbitrary function $h$, without breaking the truncated BRST symmetry. Therefore, in a formal sense, this ambiguity of the renormalization for the gauge fixing term could break the multiplicative renormalizability of the gauge fixing and the F-P ghost sector in the truncated version (\ref{eq:BRST-invariant-stochastic-action-eq2}). In Appendix B, we prove the renormalizability of the BRST invariant stochastic action principle in terms of the Ward-Takahashi identities for the 1-P-I vertices. To avoid the ambiguity in the renormalization of the gauge fixing term, we prove the renormalizability of the extended BRST invariant stochastic action (\ref{eq:BRST-invariant-stochastic-action-eq1}) instead of (\ref{eq:BRST-invariant-stochastic-action-eq2}). We obtain the renormalized stochastic action and the counter-terms as 
\bea
\label{eq:extended-counterterm-eq1} 
K_{\rm BRST} (\varphi^\a_0, g_0, \kappa_0, \xi_0) 
\equiv 
K_{\rm BRST} (Z_{\varphi^\a}^{1/2} \varphi^\a, Z_g g, Z_\kappa \kappa, Z_\xi \xi)          \ , 
\ena
where 
$
\varphi^\a \equiv (V, \varpi, c, {\bar c}, c', {\bar c}', \phi, {\bar \phi}, B, {\bar B})      
$. The subscript \lq\lq 0 \rq\rq denotes the bare quantities. 
The renormalization constants satisfy the relations 
\bea
\label{eq:relation-renormalization-constants-eq1}
Z_\kappa^{-1} Z_B^{1/2} Z_\phi^{1/2} = 1,\ 
Z_\phi^{1/2} = Z_\xi Z_V^{1/2} = Z_{\bar \phi}^{1/2} ,\ 
Z_B = Z_{\bar B} = Z_{c'}=Z_{{\bar c}'}  \ . 
\ena
The first two relations in (\ref{eq:relation-renormalization-constants-eq1}) are consequences of the non-renormalization of the stochastic gauge fixing term in (\ref{eq:BRST-invariant-stochastic-action-eq1}). 

Since the truncated version is derived from the extended version by integrating out the auxiliary fields $(\phi, {\bar \phi}, B, {\bar B})$, the two formulations are formally equivalent. Furthermore, the expectation values of local gauge invariant observables are independent of the stochastic gauge fixing functions, and F-P ghosts do not contribute to the renormalization. Therefore, the ambiguity in the renormalization of the gauge fixing term in the truncated BRST invariant formulation may be harmless for actual calculations.  In fact, the perturbative analysis in terms of the truncated version of the stochastic action in the background field method is consistent with that of the conventional approach in the one-loop approximation, which is confirmed by the explicit calculation of the $\beta$-function of the gauge coupling.\cite{Nakazawa5}

\section{Discussion} 

We have introduced the stochastic gauge fixing procedure for ${\cal N} = 1$ SYM$_4$ in the SQM approach in which the superfield formalism preserves the local gauge symmetry as well as the global supersymmetry in the underlying stochastic process in the sense of It${\bar {\rm o}}$ calculus. We have shown that the standard Faddeev-Popov probability distribution is a stationary solution of the Fokker-Planck equation, and it is conjectured to be reproduced in the equilibrium limit. This means that the contribution of the Faddeev-Popov ghost superfield in the path-integral approach is simulated by the extra self-interaction of the vector superfield in the stochastic gauge fixing term. 
The perturbative properties have been studied in terms of the BRST invariant stochastic action with regularization by dimensional reduction. In particular, the non-renormalization theorem holds, as in the conventional approach. We have also shown the perturbative renormalizability of the present formulation based on the power counting argument and the BRST symmetry of the stochastic action. It is essential for the SQM approach to preserve the manifest supersymmetry of SYM$_4$ that the supersymmetry is defined as a translation in the superspace in terms of the superfield formalism. In the SQM approach, a possible breakdown of the supersymmetry would be induced by the correlation of the noise variables. In the superfield formalism, the correlation of the noise superfield is also invariant under the translation in the superspace. Therefore, the time development with respect to the stochastic time preserves the supersymmetry, and thus the system is invariant under the translation $at$ $finite$ $stochastic$ $time$. In this respect, SQM can be applied directly to the supersymmetric models formulated in terms of the superfield formalism. By contrast, if we apply SQM to a supersymmetric model that consists of only physical component fields, we would encounter an artificial breakdown of the supersymmetry $at$ $finite$ $stochastic$ $time$, though the supersymmetry is expected to be recovered $in$ $equilibrium$. 

In this paper, we have formulated the stochastic action on the basis of the BRST symmetry. Since the interaction is non-polynomial for SYM$_4$ in the superfield formalism, it is convenient to apply the background field method (BFM) for perturbative calculations in the SQM approach\cite{Okano}. In the conventional (path-integral) BFM for SYM$_4$ without chiral matter multiplets,\cite{GSR} the chiral superfields, such as the Nakanishi-Lautrup fields and Faddeev-Popov ghosts, are defined by the chiral condition with respect to the $background$ $covariant$ derivative. This causes a non-trivial contribution of the Nakanishi-Lautrup fields. In order to cancel this contribution, we need an additional ghost, i.e., Nielsen-Kallosh-type ghost.\cite{NK} In Ref.\citen{Nakazawa5} we carried out a one-loop calculation to determine the $\beta$-function of the gauge coupling by applying BFM to the stochastic action. The one-loop $\beta$-function is identical to the known results in the path-integral approach. This indicates that the contribution of the Nielsen-Kallosh ghost can be simulated by defining the stochastic gauge fixing function in terms of the backgound covariant derivatives, instead of $D_\a$ and ${\bar D}_{\dot \a}$ in this paper. We note that the necessity of the Nielsen-Kallosh ghost is peculiar to the conventional BFM. 

One important check for the equivalence of the SQM approach to the standard path-integral approach is to evaluate anomalies in SYM$_4$\cite{CPS-PiSi}. In Ref.\citen{Nakazawa6}, we derived the anomalous Ward-Takahashi identity for the superconformal symmetry in this context with the stochastic gauge fixing procedure in the one-loop approximation. We also confirmed the one-loop equivalence of the effective equation of motion obtained in the SQM approach and the conventional one, which includes the contribution from ghosts. The equivalence in BFM is shown by expressing the ghost superpropagators in terms of the vector superpropagator with the help of Slavnov-Taylor identities and Schwinger-Dyson equations at the one-loop level. This explicit one-loop equivalence in BFM also supports the formal proof given in this paper. In the conventional approach, there is no anomaly for the local gauge symmetry, because the gauge multiplet and the ghost multiplet belong to the adjoint representation. This is also true in the SQM approach, due to the one-loop equivalence in BFM to the conventional approach. In this paper, we have considered only the vector superfield, i.e., the gauge multiplet in SYM$_4$ including F-P ghosts in the conventional sense. The extension to introduce matter fields, such as chiral matter multiplets in any representations, does not appear to present any obstacles for the case of the anomaly free combination. 

We have mainly investigated perturbative properties of SYM$_4$ in the SQM approach with the stochastic gauge fixing procedure. In order to give a precise definition of the stochastic process for SYM, it is necessary to reduce the number of degrees of freedom to a finite value for the existence of a unique equilibrium limit. The best way to carry this out may be to apply the lattice regularization. However, in the lattice regularization, the exact global supersymmetry is not manifest, though it is expected to be recovered in the continuum limit. In this respect, we may regard the naive zero volume limit of SYM at large $N$ as a regularization procedure that preserves the global supersymmetry and allows us to define the stochastic process precisely, because the finiteness of the partition function and correlation functions in this limit has been proven with finite $N$.\cite{AW} In addition, though the fermion integral in SYM yields a Pfaffian, it is positive semi-definite in the four-dimensional case. Therefore, the zero volume limit of SYM$_4$ in the superfield formalism, which is expected to recover the planar limit of the continuum SYM$_4$, may provide a toy model to study the stochastic process with exact global supersymmetry in the context of the SQM approach. 

We finally comment on some difficulties in the application of SQM to the constructive definition of superstring theories in terms of supersymmetric large $N$ reduced models.\cite{M} As mentioned above, without auxiliary fields in the superfield formalism, we would encounter an artificial breakdown of the supersymmetry $at$ $finite$ $stochastic$ $time$. 
Although we are able to expect that the supersymmetry is recovered in the limit of infinite stochastic time for the cases of SYM$_{10}$ and its zero volume limit,\cite{Nakazawa1} there may be no merit of considering a naive application of SQM to such models, because the Fokker-Planck hamiltonian, which is Lorentz invariant by construction, does not commute with the generator of the supersymmetry. One way to avoid this problem is to integrate out the fermion fields. However, in contrast to the case for SYM$_4$, the Pfaffian is complex in the ten-dimensional case, and this may cause another difficulty in attempting to define the stochastic process in terms of the Langevin equation. 
These problems remain to be solved, and we hope that the present analysis of SYM$_4$ in the superfield formalism provides an insight for future application of SQM to supersymmetric models, such as SYM in the Wess-Zumino gauge, which consist of only physical component fields.

\section*{Acknowledgements}

This work was completed at KEK. The author would like to thank all the members of the theory group at KEK for their hospitality. He also wishes to thank H. Kawai and T. Tada for encouragement.


\appendix


\section{Conventions for SYM$_4$ in the SQM Approach}\label{sec:A}

In this appendix, we summarize the superfield formalism of 
SYM$_4$ in the SQM approach. We use a convention closely related to that of Ref.\citen{WB}. In particular, the reality condition on the vector superfield $V^\dagger = V$ is defined in the Minkowski space-time with the signature $(-,+,+,+)$. In order to assure the convergence of the superfield Langevin equation, we consider the Wick rotation to the Euclidean space-time by $i x^0 \equiv x^4$. In this case, the generators of the $SL(2,{\bf C})$ are mapped to those for $SO(4) = SU(2)\times SU(2)$ by preserving the relation $\sigma^m \pa_m = \sigma^m_{(\rm E)} \pa_m^{(\rm E)}$. It is not necessary to change the algebraic rule for the $D$-algebra, which is obtained by replacing $\eta^{mn} \rightarrow \delta^{mn}$. For example, $\square = \pa^2_1 + \pa^2_2 + \pa^2_3 + \pa^2_4$ is understood.\cite{Kalivas} 
In the Euclidean space-time, however, since $\theta_\a^\dagger \neq {\bar \theta}_{\dot \a}$, the superfield formulation needs a doubled superspace, $(x,\theta_\a,{\bar \theta}_{\dot \a}) \equiv S_1$ and $(x^\dagger, {\bar \theta}^\dagger_{\dot \a}, \theta_\a^\dagger) \equiv S_2$, where $\theta_\a$ and ${\bar \theta}_{\dot \a}^\dagger$ are two independent spinor coordinates.\cite{LN} Thus, the number of degrees of freedom of the vector multiplet are also doubled, and hence $\lambda_\a^\dagger \neq {\bar \lambda}_{\dot \a}$, $v_m^\dagger \neq v^m$, etc. In other words, suppose that the vector superfield is defined on $S_1$, i.e., $V \in S_1$, the hermitian conjugation is defined on $S_2$, i.e., $V^\dagger \in S_2$. Then the hermiticity of the vector superfield is defined by the self-conjugation in the sense of Ostewarlder-Schrader by identifying the doubled superspace, $S_1$ and $S_2$, with the simultaneous time reversal operation.  The identification of the doubled fermionic component fields is defined by the Majorana condition, $\lambda_\a^\dagger = {\bar \lambda}_{\dot \a}$, $by$ $continuing$ $back$ $to$ $the$ $Minkowski$ $space$-$time$. This prescription is also applied to the chiral superfield. If we difine a chiral superfield $\Psi$ and an anti-chiral superfield ${\bar \Psi}$ on $S_1$, i.e., ${\bar D}_{\dot \a} \Psi = D_\a {\bar \Psi} = 0$, the hermitian conjugation is defined on $S_2$, i.e., $\Psi^\dagger \in S_2$. Then $\Psi^\dagger$ can be identified with ${\bar \Psi} \in S_1$ in the sense of the O-S conjugation. Therefore, if we study the hermiticity of the vector superfield in the Euclidean superspace in terms of this doubled formulation, we obtain no physically distinct conclusions in comparison with the hermiticity condition of the vector superfield in the Minkowski space-time. In this respect, it is safe to use the superfield formulation of the vector superfield in the Minkowski space-time, even after the Wick rotation, if we keep in mind that the hermitian conjugation is defined in the sense of Ostewarlder-Schrader.

The vector superfield $V$ is $SU(N)$ algebra valued: $V \in su(N)$ ; $V = V^a t_a$, $[ t_a, t_b] = if_{abc} t_c$ and ${\rm Tr}( t_a t_b ) = k \delta_{ab}$. For the convenience, ${t^a}$ is assumed to be in the fundamental representaion.  
The action for SYM$_4$ is defined by 
\bea
\label{eq:4Daction-A-eq1}
S 
 =  
- \int d^4x d^2\theta d^2{\bar \theta} \displaystyle{\frac{1}{4k g^2}} {\rm Tr}    \Big(
W^\alpha W_\alpha \delta^2 ( {\bar \theta} ) + {\bar W}_{\dot \alpha}{\bar W}^{\dot \alpha} \delta^2 ( \theta )
\Big)           \  . 
\ena
Here 
\bea
\label{eq:Gluino-field-A-eq1}
W_\alpha 
 = &
- \displaystyle{\frac{1}{8}} {\bar D}^2 e^{-2gV}D_\alpha e^{2gV} 
     \ , \qquad 
{\bar W}_{\dot \alpha} 
=  
 \displaystyle{\frac{1}{8}} D^2 e^{2gV}{\bar D}_{\dot \alpha} e^{-2gV}            \  .
\ena

The analogues of the Lie derivatives and the Haar measure defined on a group manifold are introduced as follows. The differential operator 
${\hat {\cal E}}^{(L,\ R)}_a(z)$ is defined by
\bea
\label{eq:analogue-Lie-derivative-A-eq1}
{\hat {\cal E}}^{(L)}_a (z) e^{2g V (z')} 
& = & 
t_a e^{2g V (z')}\delta^8 ( z -z' )    \ , \nn
{\hat {\cal E}}^{(R)}_a (z) e^{2g V (z')} 
& = & 
e^{2g V (z')}t_a \delta^8 ( z -z' )    \ . 
\ena 
Here 
\bea
\label{eq:analogue-Lie-derivative-A-eq2}
{\hat {\cal E}}^{(L)}_a (z) 
& \equiv & 
\displaystyle{\frac{1}{2g}}
 L_a^{\ b} (z) \displaystyle{\frac{\delta\ \ }{\delta V^b (z)}}   \ , \nn 
 {\hat {\cal E}}^{(R)}_a (z) 
& \equiv & 
\displaystyle{\frac{1}{2g}}
 L_{\ a}^b (z) \displaystyle{\frac{\delta\ \ }{\delta V^b (z)}}   \ , \nn 
L_a^{\ b} (z)
& \equiv & 
\displaystyle{\frac{1}{k}} {\rm Tr}\Big( 
t_a \cdot \displaystyle{\frac{2g L_{V(z)}}{1 - e^{-2g L_{V(z)}}}} t^b  
\Big)                                                             \ , 
\ena 
where $z \equiv ( x, \theta, {\bar \theta} )$, 
$ \delta^8 ( z -z' ) \equiv \delta^4 ( x -x' )\delta^2 ( \theta - \theta' )\delta^2 ( {\bar \theta} - {\bar \theta}' )$, 
 $L_V X \equiv [ V, X]$,\cite{GGRS} 
 and 
 $L^b_{\ a} = L_a^{\ b\ \dagger}$. 
These differential operators satisfy the algebra
\bea
\label{eq:analogue-Lie-derivative-A-eq3}
 \Big[ {\hat {\cal E}}^{(L)}_a (z), {\hat {\cal E}}^{(L)}_b (z') \Big] 
& = & 
-i f_{abc} \delta^8 (z - z') {\hat {\cal E}}^{(L)}_c (z)     \ , \nn
 \Big[ {\hat {\cal E}}^{(R)}_a (z), {\hat {\cal E}}^{(R)}_b (z') \Big] 
& = & 
i f_{abc} \delta^8 (z - z') {\hat {\cal E}}^{(R)}_c (z)     \ , 
\ena
which imply 
\bea
\label{eq:analogue-Lie-derivative-A-eq4}
& {} & 
L_a^{\ c} \pa_c L_b^{\ d} - L_b^{\ c} \pa_c L_a^{\ d} 
 =  -2i g f_{abc} L_c^{\ d}              \ , \nn
& {} & 
L^c_{\ a} \pa_c L^d_{\ b} - L^c_{\ b} \pa_c L^d_{\ a} 
 =  2i g f_{abc} L^d_{\ c}               \ , \nn
& {} & 
L_a^{\ c}\pa_c L^d_{\ b} - L^c_{\ b}\pa_c L_a^{\ d} = 0 \ .  
\ena
We also define a measure by introducing the metric 
$ G^{ab} (z) \equiv L_c^{\ a} (z) L_c^{\ b} (z) 
=  L_{\ c}^a (z) L_{\ c}^b (z)$ 
and 
$ G_{ab} (z) \equiv K_a^{\ c} (z) K_b^{\ c} (z)$, 
where 
$K_a^{\ c} (z) L_c^{\ b} (z) 
 =  
L_a^{\ c} (z) K_c^{\ b} (z) = \delta_a^{\ b} $. 
By the metric, the integration measure is defined by 
$\sqrt{G} {\cal D}V  $.

The superfield Langevin equation for SYM$_4$ is given by  
\bea
\label{eq:SYM-Langevin-A-eq1}
\displaystyle{\frac{1}{2g}}
( \Delta e^{2gV} )e^{-2gV} ( \t, z )
 = 
- \beta \Delta \t 2g {\hat {\cal E}}^{(L)} ( \t, z ) S + \Delta w ( \t, z )            \ .
\ena
Here 
${\hat {\cal E}}^{(L)} ( \t, z ) = t_a {\hat {\cal E}}^{(L)}_a ( \t, z ) $. 
 The  equations of motion are defined by 
$
{\cal D}{\cal W} \equiv \Big\{ {\cal D}^\a, W_\a \Big\}  ,\ 
$
$
{\overline {\cal D}}{\overline {\cal W}} \equiv \Big\{ {\bar {\cal D}}_{\dot \a}, {\bar W}^{\dot \a} \Big\}  ,\ 
$ where 
$
{\cal D}_\a \equiv e^{-2gV} D_\a e^{2gV} 
$ and 
$
{\bar {\cal D}}_{\dot \a} \equiv e^{2gV} {\bar D}_{\dot \a} e^{-2gV} 
$. 
Then (\ref{eq:SYM-Langevin-A-eq1}) becomes (\ref{eq:SYM-Langevin-eq1}). 
Since the reality condition implies 
$e^{2g L_V} {\cal D}{\cal W} = {\overline {\cal D}}{\overline {\cal W}}$,\cite{GSR} we may discard one of them in (\ref{eq:SYM-Langevin-eq1}). 

The probability distribution $P( \t, e^{2gV} )$ is defined in terms of the expectation value of an arbitrary observable 
$F [ e^{2gV} ]$ as 
\bea
\label{eq:SYM-probability-distribution-A-eq1}
\langle F [ e^{2gV} ] ( \t ) \rangle 
\equiv \int\!\!\! F [ e^{2gV} ] P( \t , e^{2gV} ) \sqrt{G} {\cal D}V        \ . 
\ena
Here, the expectation value on the l.h.s. is evaluated, using the solution of the Langevin equation (\ref{eq:SYM-Langevin-eq1}),  by accounting for the correlations of all the noise superfields 
$ 
\{ \Delta\omega | \Delta\omega (\t') \le \t - \Delta \t \} . 
$

The Langevin equation (\ref{eq:SYM-Langevin-A-eq1}) (i.e., (\ref{eq:SYM-Langevin-eq1})) is converted to that for the vector superfield $V$ which appears in a geometrical expression 
\bea
\label{eq:SYM-Langevin-A-eq2}
\Delta V^a ( \t, z )
& = & 
- \beta \Delta \t G^{ab} ( \t, z ) \displaystyle{\frac{\delta S}{\delta V^b (z)}} 
+ \Delta_w \Xi^a ( \t, z )    \ , \nn
\langle 
\Delta_w \Xi^a ( \t, z ) \Delta_w \Xi^b ( \t, z' ) 
\rangle_{\Delta w_\t} 
& = & 
2 \beta \Delta \t G^{ab} ( \t, z )  \delta^8 ( z - z' )   \ . 
\ena
Here, $\Delta_w \Xi^a$ is a collective noise superfield defined as  
$ \Delta_w \Xi^a ( \t, z ) \equiv \Delta w^b ( \t, z ) L_b^{\ a} ( \t, z ) $. 
In the matrix representation, (\ref{eq:SYM-Langevin-A-eq2}) becomes (\ref{eq:SYM-Langevin-eq2}). In terms of the It${\bar {\rm o}}$ calculus, the noise variable is uncorrelated with the equal  stochastic-time dynamical variables. This means that the r.h.s. of the collective noise correlation (\ref{eq:SYM-Langevin-A-eq2}) is $not$ the expectation value. 

The vector superfield must remain real under the time evolution described by (\ref{eq:SYM-Langevin-A-eq1}), i.e.,  
$V( \t + \Delta \t )^\dagger = V( \t + \Delta \t )$. 
To satisfy this constraint, we impose the hermitian conjugation of the noise superfield as 
\bea
\label{eq:noise-hermiticity-A-eq1}
\Delta w( \t )^\dagger 
 =   e^{-2g L_V} \Delta w ( \t )    \  . 
\ena
 This is consistent with the transformation property of $\Delta w$ given in (\ref{eq:noise-transformation}). Although $\Delta w$ is not a vector superfield, 
 it follows from (\ref{eq:noise-hermiticity-A-eq1}), 
\bea
\label{eq:noise-reality-A-eq1}
{ \Delta_w \Xi^a }^\dagger 
& = & 
{ \Delta w^b }^\dagger { L_b^{\ a} }^\dagger         \  ,    \nn 
& = & 
\displaystyle{\frac{1}{{\rm k}}} {\rm Tr}\Big( 
\Delta w^\dagger \cdot \displaystyle{\frac{2g L_{V(z)}}{e^{2g L_{V(z)}} - 1}} t^b    \Big)   
 = 
 \Delta_w \Xi^a      \ . 
\ena
Thus, the collective noise superfield is a vector superfield. 
This ensures that the time development described by (\ref{eq:SYM-Langevin-A-eq2}) (i.e., (\ref{eq:SYM-Langevin-eq2})) preserves the hermiticity of the vector superfield. 

The Fokker-Planck equation corresponding to (\ref{eq:SYM-Langevin-A-eq2}) is expressed as 
\bea 
\label{eq:SYM-Fokker-Planck-A-eq2}
\displaystyle{\frac{\pa}{\pa \t}} P( \t, V ) 
= 
\beta 
\int\!\!\! d^8z \displaystyle{\frac{\delta\ }{\delta V^a (z)}}  \Big\{ 
G^{ab}( \t, z ) \Big( 
\displaystyle{\frac{\delta\ }{\delta V^b (z)}} + 
 \displaystyle{\frac{\delta S}{\delta V^b (z)}} 
 \Big)
 P( \t, V )  \Big\}         \ . 
\ena

The hermitian conjugation of the noise components is uniquely determined from the hermiticity assignment (\ref{eq:noise-hermiticity-A-eq1}). Here we summarize the results in the \lq\lq almost Wess-Zumino gauge \rq\rq\ in which we can expand  
\bea
\label{eq:noise-almost-WZ-A-eq1}
\Delta w^\dagger 
& = & 
\Delta w - 2g [ V|_{\rm WZ},\ \Delta w ] + 2g^2 [ V|_{\rm WZ},\ [ V|_{\rm WZ},\ \Delta w ]]  \ , \nn
V|_{\rm WZ} 
& \equiv & 
- \theta\sigma^m {\bar \theta} v_m (x)        
 + 
i \theta^2 {\bar \theta}  {\bar \lambda} (x)         
- i {\bar \theta}^2 \theta \lambda (x) 
+ \displaystyle{\frac{1}{2}}\theta^2 {\bar \theta}^2  D (x) \ . 
\ena 
This relation determines the imaginary part of the complex noise $\Delta w$. For example, for the vector component of $\Delta w$, (\ref{eq:noise-almost-WZ-A-eq1}) reads  
$
\Delta w |^\dagger_{\theta\sigma^m {\bar \theta}} 
= \Delta w |_{\theta\sigma^m {\bar \theta}} 
+ 2g[ v_m,\ \Delta w |\ ]  
$. Here $\Delta w |$ is the lowest component of $\Delta w$, and $\Delta w |^\dagger = \Delta w |$. 
This implies that the imaginary part of the vector component 
$\Delta w |_{\theta\sigma^m {\bar \theta}} $ 
is given by 
$- g[ v_m,\ \Delta w |\ ] $. 
By contrast, 
the collective noise superfield $\Delta_w \Xi$ has a manifestly hermitian expression, 
\bea
\label{eq:noise-almost-WZ-A-eq2}
\Delta_w \Xi 
& = & \displaystyle{\frac{1}{2}}\Delta w + 
\displaystyle{\frac{1}{2}}\Delta w^\dagger + \displaystyle{\frac{g^2}{6}} [ V|_{\rm WZ},\ [ V|_{\rm WZ},\  \Delta w + \Delta w^\dagger ]]      \ . 
\ena

Since $\Delta_w \Xi$ is a vector superfield, it can be expanded into its components in a standard manner: 
$\Delta_w \Xi = \Delta_w \Xi | + ... - \theta\sigma^m {\bar \theta} ( \Delta_w \Xi |_{\theta\sigma^m {\bar \theta}}) + ... $ . The correlation of each component of the collective noise is uniquely determined from (\ref{eq:SYM-Langevin-eq1}). The details of the component expansion are given in Ref.\citen{Nakazawa1}. For example, for the vector components, the correlation in (\ref{eq:SYM-Langevin-A-eq2}) reads 
\bea
\label{eq:noise-almost-WZ-A-eq3}
\langle 
( \Delta_w \Xi |_{\theta\sigma^m {\bar \theta}}) (\t, x )( \Delta_w \Xi |_{\theta\sigma^n {\bar \theta}}) (\t, y ) \rangle 
= -4 \beta \Delta\t \delta_{mn} \delta^4( x-x' )  
        \ .
\ena
In the superfield formalism, the precise meaning of a superfield expression is defined by its component expressions. In this respect, though the equilibrium distribution is independent of the value of the parameter $\beta$, we need to choose $\beta = - {1\over2}$ in the superfield Langevin equation and the noise correlation in (\ref{eq:SYM-Langevin-eq1}) to provide a proper normalization in their component expressions with the conventions of Ref.\citen{WB} in the Euclidean space-time. In this note, we introduce a scaling parameter $\kappa$ for the renormalization procedure. Specifically, we define 
$\beta \equiv - \displaystyle{\frac{1}{2\kappa}}$, as we define the BRST symmetry (\ref{eq:stochastic-BRST-tr.-eq1}) and the BRST invariant stochastic action (\ref{eq:BRST-invariant-stochastic-action-eq1}). 


\section{Renormalizability of SYM$_4$ in the SQM Approach}\label{sec:B}

In this appendix, we demonstrate the renormalizability of the SQM approach in terms of the BRST invariant stochastic action principle formulated in this paper. For the YM$_4$ case, the renormalizability was proved formally in Ref.\citen{ZZ}. That proof is based on the Ward-Takahashi identities derived from the truncated BRST symmetry, which corresponds to (\ref{eq:stochastic-BRST-tr.-eq2}) for the SYM$_4$ case. The truncated BRST invariant formulation may be convenient, because of the reduced number of the auxiliary fields. However, the local identities for the changes of the integration variables 
 $c' \rightarrow c' + \delta c'$ and ${\bar c}' \rightarrow {\bar c}' + \delta {\bar c}'$ include composite operators that are not identified with the variation under the truncated BRST transformation (\ref{eq:stochastic-BRST-tr.-eq2}). This fact complicates the application of the standard procedure to determine the divergent terms for the stochastic gauge fixing and the FP ghost sector in terms of the Ward-Takahashi identity. 
 
Instead of employing the truncated BRST formulation, we demonstrate the renormalizability in the SQM approach in terms of the extended (or \lq\lq 5-dimensional \rq\rq) BRST symmetry (\ref{eq:stochastic-BRST-tr.-eq1}) for the extended BRST invariant stochastic action (\ref{eq:BRST-invariant-stochastic-action-eq1}). Although, this includes extra auxiliary superfields ($\phi$, ${\bar \phi}$) as the \lq\lq 5-th component \rq\rq of the gauge field and the Nakanishi-Lautrup fields, ($B$, ${\bar B}$), provided that we do not evaluate correlation functions that include these auxiliary fields explicitly, the extended BRST invariant stochastic action (\ref{eq:BRST-invariant-stochastic-action-eq1}) is equivalent to the truncated one (\ref{eq:BRST-invariant-stochastic-action-eq2}). Furthermore, the expectation values of local gauge invariant observables as functionals of $V$ are independent of the stochastic gauge fixing functions $\phi$ and ${\bar \phi}$, and the F-P ghost loop effect gives no contribution to the renormalization in either (\ref{eq:BRST-invariant-stochastic-action-eq1}) or (\ref{eq:BRST-invariant-stochastic-action-eq2}), due to the retarded (or causal) property of their superpropagators. The decoupling of FP ghosts means that we may discard the FP ghost sector for the calculation of observables. 

Let us consider the generating functional of the stochastic Feynman diagrams for (\ref{eq:BRST-invariant-stochastic-action-eq1}) by introducing the source terms 
\bea
\label{eq:K-external-sources-B-eq1}
K_{\rm ex} 
& = & 
\int_0 \!\! \Big( J_V^a V^a + J_\varpi^a \varpi^a \Big) 
+ 
\int_{+} \!\!\! \Big( J_\phi^a \phi^a + J_c^a c^a + J_{c'}^a c^{' a} + J_B^a B^a \Big)     \nn
& {} & \quad 
+ 
\int_{-} \!\!\! \Big( J_{\bar \phi}{\bar \phi}^a + J_{\bar c}^a{\bar c}^a 
+ J_{{\bar c}^{' a}} {\bar c}^{' a}  +  J_{\bar B}^a {\bar B}^a  \Big) \nn
& \equiv & 
\varint \sum_\a J_{\varphi^\a} \cdot \varphi^\a \ , 
\ena
where $\int_0 \equiv \int\!\! d^8z d\t$, $\int_{+} \equiv \int\!\! d^6z \d\t$ and $\int_{-} \equiv \int\!\! d^6{\bar z}d\t$. Here, $\varint$ denotes the appropriate measure for the integrations with respect to the superspace coordinates and the stochastic-time. The dot ( \lq\lq{ $\cdot$ }\rq\rq ) denotes the sum with respect to the group indices. We also introduce the external source terms 
\bea
\label{eq:K-external-sources-B-eq2}
K_{\rm ex}^{\rm BRST} 
& = & 
\int_0 \!\!\! \Big( I_V^a (sV^a) + I_\varpi^a (s\varpi^a) \Big) + 
\int_{+} \!\!\! \Big( I_\phi^a (s\phi^a) + I_c^a (sc^a) + I_{c'}^a (sc^{' a}) \Big) \nn
& {} & \quad 
+ 
\int_{-} \!\!\! \Big( I_{\bar \phi} (s{\bar \phi}^a) + I_{\bar c}^a (s{\bar c}^a) 
+ I_{{\bar c}^{' a}} (s{\bar c}^{' a}) \Big) \nn
& \equiv & 
\varint \sum_\a I_{\varphi^\a} \cdot (s\varphi^\a) \ . 
\ena
Here we note that $\delta_{\rm BRST} \varphi^\a \equiv (s\varphi)^\a$, as defined in (\ref{eq:stochastic-BRST-tr.-eq1}) and (\ref{eq:redefined-BTST-eq1}). 
The external source $I_{\varphi^\a}$ has the same chirality as the original superfield $\varphi^\a$. 
The transformational and dimensional properties of the fields and the external sources are listed in Table I. (In this table, the chirality + ($-$) is defined by ${\bar D} \varphi = 0$ ($D \varphi = 0$).) 
%
%

\begin{table}
 \caption{Dimensional properties of fields and sources}
 \begin{center}
 \begin{tabular}{ccccccccccc} \hline  \hline
  ${\rm field}$      & $V$ & $\varpi$ & $c$ & ${\bar c}$ & $c'$ & ${\bar c}'$     & $\phi$ & ${\bar \phi}$ & $B$ & ${\bar B}$                                                      \\ \hline
  ${\rm dim.}$       & 0   &   2      &   0 &   0        &  3   &  3              &  2     &  2            & 3   &   3                                                             \\ \hline   
  $\#{\rm ghost}$    & 0   & 0        & 1   & 1          &  $-1$ & $-1$   
  &  0     &  0            & 0   & 0                                                             \\ \hline    
  ${\rm chirality}$  & 0   & 0        & +   & $-$        &  +    & $-$   
  &  +     &  $-$          & +   & $-$                                                           \\ \hline 
 \end{tabular}
 \end{center}
 \label{table:1}
\end{table}

\begin{table}  
 \begin{center}
 \begin{tabular}{ccccccccccc} \hline  \hline 
  ${\rm source}$     & $J_V$ & $J_\varpi$ & $J_c$ & $J_{\bar c}$ & $J_{c'}$       & $J_{{\bar c}'}$     & $J_\phi$ & $J_{\bar \phi}$ & $J_B$ & $J_{\bar B}$                                                    \\ \hline
  ${\rm dim.}$       & 4     &   2        &   5   &   5          &  2             &  2                &  3       &  3              & 2     &   2                                                             \\ \hline   
  $\#{\rm ghost}$    & 0     & 0          & $-1$  & $-1$         &  1             &  1                &  0       &  0              & 0     & 0                                                               \\ \hline    
  ${\rm chirality}$  & 0     & 0          & +     & $-$          &  +             & $-$               &  +       &  $-$            & +     & $-$                                                              \\ \hline 
 \end{tabular}
 \end{center}
\end{table}

\begin{table}
 \begin{center}
 \begin{tabular}{ccccccccc} \hline \hline 
  ${\rm source}$     & $I_V$ & $I_\varpi$ & $I_c$ & $I_{\bar c}$ & $I_{c'}$       & $I_{{\bar c}'}$     & $I_\phi$ & $I_{\bar \phi}$                                                     \\ \hline
  ${\rm dim.}$       & 4     &   2        &   5   &   5          &  2             &  2                  &  3       &  3                                                                      \\ \hline   
  $\#{\rm ghost}$    & $-1$  & $-1$       & $-2$  & $-2$         &  0             &  0                  &  $-1$    &  $-1$                                                                      \\ \hline    
  ${\rm chirality}$  & 0     & 0          & +     & $-$          &  +             & $-$                 &  +       &  $-$                                                                    \\ \hline 
 \end{tabular}
 \end{center}
\end{table}

%

The connected stochastic Feynman diagrams are generated by $W(J, I)$ as 
\bea
\label{eq:W-functional-B-eq1}
Z (J, I) = \int \!\!\! d\mu e^{K_{\rm tot}} \equiv e^{W(J, I)} \ , \quad 
K_{\rm tot} = K_{\rm BRST} + K_{\rm ex} + K_{\rm ex}^{\rm BRST} \ , 
\ena
where $d\mu = {\cal D}V{\cal D}\varpi{\cal D}c{\cal D}{\bar c}{\cal D}c'{\cal D}{\bar c}'{\cal D}\phi{\cal D}{\bar \phi}{\cal D}B{\cal D}{\bar B}$.
The effective stochastic action, i.e., the generator of the 1-P-I vertices in the stochastic Feynman diagrams, is defined by 
\bea
\label{eq:Gamma-functional-B-eq1}
\Gamma({\bar \varphi}, I) 
& = & 
W(J, I) - \varint \sum_\a J_{\varphi^a} {\bar \varphi}^\a    \ , \nn 
{\bar \varphi}^\a   
& = & 
\displaystyle{\frac{\delta W}{\delta J_{\varphi^\a}}} , \ 
J_{\varphi^\a} = - (-1)^{\varphi^\a} \displaystyle{\frac{\delta \Gamma}{\delta {\bar \varphi}^\a}}        \ . 
\ena
Hereafter, we omit \lq\lq $-$ \rq\rq for the vacuum expectation values of the superfield 
$\varphi^\a \equiv (V, \varpi, c, {\bar c}, c', {\bar c}', \phi, {\bar \phi}, B, {\bar B})$ so that these are not confused with their hermitian conjugates. 

In terms of the effective stochastic action, the Ward-Takahashi identity for the extended BRST symmetry (\ref{eq:stochastic-BRST-tr.-eq1}) is 
\bea
\label{eq:ST-identity-B-eq1}
\varint \sum_\a \displaystyle{\frac{\delta \Gamma}{\delta \varphi^\a}} 
\displaystyle{\frac{\delta \Gamma}{\delta I_{\varphi^\a}}} 
= \Gamma \ast \Gamma = 0 \ . 
\ena
In order to show the non-renormalization theorem for the stochastic gauge fixing term, we derive local identities by changing the integration variable as 
$c' \rightarrow c' + \delta c'$, with arbitrary $\delta c'$.  A similar identity is also derived for ${\bar c}' \rightarrow {\bar c}' + \delta{\bar c}'$. In terms of $\Gamma$, these local identities are expressed as 
\bea
\label{eq:Gamma-C'-identity-B-eq1}
& {} & 
\displaystyle{\frac{1}{2\kappa}} \left( \displaystyle{\frac{\delta \Gamma}{\delta I_\phi}} - \displaystyle{\frac{i}{4}}\xi {\bar D}^2D^2 \displaystyle{\frac{\delta \Gamma}{\delta I_V}}\right) - \displaystyle{\frac{\delta \Gamma}{\delta c'}} = 0 \ ,     \nn 
& {} & 
\displaystyle{\frac{1}{2\kappa}} \left( \displaystyle{\frac{\delta \Gamma}{\delta I_{\bar \phi}}} + \displaystyle{\frac{i}{4}}\xi D^2{\bar D}^2 \displaystyle{\frac{\delta \Gamma}{\delta I_V}}\right) - \displaystyle{\frac{\delta \Gamma}{\delta {\bar c}'}} = 0 \ .   
\ena
Similarly, for 
$B \rightarrow B + \delta B$ and $
{\bar B} \rightarrow {\bar B} + \delta{\bar B}$, we obtain 
\bea
\label{eq:Gamma-B-identity-B-eq1}
- \displaystyle{\frac{i}{2\kappa}}\Phi - \displaystyle{\frac{\delta \Gamma}{\delta B}} + i I_{c'} = 0   ,    \nn 
- \displaystyle{\frac{i}{2\kappa}}{\bar \Phi} - \displaystyle{\frac{\delta \Gamma}{\delta {\bar B}}} + i I_{{\bar c}'} = 0 \ . 
\ena
From (\ref{eq:Gamma-B-identity-B-eq1}), the ($B,{\bar B}$)-dependence of $\Gamma$ is determined as 
\bea
\label{eq:Gamma-B-dependence-B-eq1}
\Gamma 
= {\tilde \Gamma} + \int_{+} \!\!\! B \Bigl( - \displaystyle{\frac{i}{2\kappa}}\Phi + i I_{c'} \Bigr) + 
\int_{-} \!\!\! {\bar B} \Bigl( - \displaystyle{\frac{i}{2\kappa}}{\bar \Phi} + i I_{{\bar c}'} \Bigr)            \  , 
\ena
where ${\tilde \Gamma}$ does not depend on $B$, ${\bar B}$, $I_{c'}$ or $I_{{\bar c}'}$. The non-renormalization of the terms which depend on the Nakanishi-Lautrup fields in $\Gamma$ is consistent with the power counting argument. 
Since $\Gamma - {\tilde \Gamma}$ does not include $c'$, ${\bar c}'$, $I_V$, $I_\phi$ or $I_{\bar \phi}$, ${\tilde \Gamma}$ satisfies the local identities (\ref{eq:Gamma-C'-identity-B-eq1}). By substituting (\ref{eq:Gamma-B-dependence-B-eq1}) into (\ref{eq:ST-identity-B-eq1}), we obtain 
\bea
\label{eq:ST-identity-B-eq2}
\varint \sum_\a \displaystyle{\frac{\delta {\tilde \Gamma}}{\delta \varphi^\a}}
\displaystyle{\frac{\delta {\tilde \Gamma}}{\delta I_{\varphi^\a}}} 
= 0                \  ,
\ena
where 
$\varphi^\a \equiv (V, \varpi, c, {\bar c}, c', {\bar c}', \phi, {\bar \phi})$. 
Let ${\tilde \Gamma}_{\rm div}$ and ${\tilde \Gamma}_{\rm R}$ be the divergent part and the finite part of the effective stochastic action; i.e., ${\tilde \Gamma} \equiv {\tilde \Gamma}_{\rm R} + {\tilde \Gamma}_{\rm div}$. We assume that the loop expansion gives ${\tilde \Gamma}_{\rm div} = \sum_{\ell = 1} (\hbar)^\ell {\tilde \Gamma}^{(\ell)}_{\rm div}$ and determine ${\tilde \Gamma}^{(1)}_{\rm div}$ from the Ward-Takahashi identity. In the following, we omit the superscript \lq\lq (1) \rq\rq. The local identity for ${\tilde \Gamma}_{\rm div}$ indicates that ${\tilde \Gamma}_{\rm div}$ takes the form 
\bea
\label{eq:tilde-Gamma-div-B-eq1}
{\tilde \Gamma}_{\rm div} 
& = & 
{\tilde \Gamma}'_{\rm div} + {\tilde \Gamma}_{\rm div}^{\rm FP} + {\tilde \Gamma}_{\rm div}^{({\rm ex})}     \ , \nn
{\tilde \Gamma}_{\rm div}^{\rm FP} 
& = & 
\displaystyle{\frac{1}{2\kappa}}\int_{+} \!\!\! c' (s'\Phi) +  
\displaystyle{\frac{1}{2\kappa}}\int_{-} \!\!\! {\bar c}' (s'{\bar \Phi}) \ , \nn 
{\tilde \Gamma}_{\rm div}^{({\rm ex})} 
& = & 
\varint \sum_\a I_{\varphi^\a} (s' \varphi^\a)   \ , 
\ena
where $(s'\varphi^\a)$ has the same dimensional properties as $(s\varphi^\a)$ given in Table I. ${\tilde \Gamma}'_{\rm div}$ is a functional of ($V, \varpi, \phi, {\bar \phi}$). 
Here, in deriving (\ref{eq:tilde-Gamma-div-B-eq1}), we have used the fact that the dependence of the divergent terms in ${\tilde \Gamma}_{\rm div}$ on the external sources $I_{\varphi^\a}$ is at most linear. From the dimensional analysis, one might suspect that there exist terms quadratic with respect to, say, 
$I_\varpi^a S_a^{\ b} c^b$ and $I_\varpi^a T_a^{\ b}{\bar c}^b$, where $S_a^{\ b}$ and $T_a^{\ b}$ are arbitrary functions of $V$. As we have already explained, because of the retarded (or causal) property of the superpropagator $\langle {V\varpi} \rangle$ with respect to the stochastic-time, a loop which consists of only $\langle {V\varpi} \rangle$ as the internal lines does not contribute to the stochastic effective action. Therefore, the appearance of a term such as $I_\varpi^a S_a^{\ b} c^b I_\varpi^c T_c^{\ d}{\bar c}^d$ is excluded. Applying the same reasoning and the power counting argument, it can be shown that ${\tilde \Gamma}'_{\rm div}$ is linear or quadratic with respect to the canonical conjugate momentum $\varpi$. In particular, there are no $\varpi$-independent terms in ${\tilde \Gamma}'_{\rm div}$ in (\ref{eq:tilde-Gamma-div-B-eq1}). This is an essential point for the renormalizability of the SQM approach in terms of the BRST invariant framework with the stochastic action principle. 

The same conclusion can be derived from the so-called $stochastic$ Ward-Takahashi identity. To derive this identity, we consider the generating functional $Z$ (or $W$) in (\ref{eq:W-functional-B-eq1}) under the condition 
$J_V = J_\phi = J_{\bar \phi} = I_V = I_\varpi = I_\phi = I_{\bar \phi} = 0$. This condition is required to perform the integration with respect to $\varpi$ without producing the non-trivial Jacobian factor. Integrating out $\varpi$ by using the same trick as in (\ref{eq:unity-eq1}), we obtain 
\bea
\label{eq:stochastic-WT-identity-eq1}
Z|_{J_V = J_\phi = J_{\bar \phi} = I_V = I_\varpi = I_\phi = I_{\bar \phi} = 0}  = {\rm independent\ of} J_\varpi   \ . 
\ena
This means 
\bea
\label{eq:stochastic-WT-identity-eq2}
\varpi \equiv 
\displaystyle{\frac{\delta G}{\delta J_\varpi}} = 0   , \quad 
{\rm for} \ J_V = J_\phi = J_{\bar \phi} = 0 \ {\rm and} \ 
I_V=I_\varpi=I_\phi=I_{\bar \phi}=0  \ .  
\ena
Therefore, in terms of the effective stochastic action, we obtain the $stochastic$ Ward-Takahashi identity
\bea
\label{eq:stochastic-WT-identity-eq3}
\displaystyle{\frac{\delta \Gamma}{\delta V}} 
= 
\displaystyle{\frac{\delta \Gamma}{\delta \phi}} 
= 
\displaystyle{\frac{\delta \Gamma}{\delta {\bar \phi}}} = 0  , \quad 
{\rm for} \ \varpi =0 \ {\rm and} \ I_V = I_\varpi = I_\phi = I_{\bar \phi}= 0 \ . 
\ena
This implies that there is no $\varpi$-independent term in ${\tilde \Gamma}'_{\rm div}$. 

Let us decompose the Ward-Takahashi identity (\ref{eq:ST-identity-B-eq2}) into two parts. 
To do this, we assume the most general form of $(s'\varphi^\a)$ for ${\tilde \Gamma}^{\rm ex}_{\rm div}$ in (\ref{eq:tilde-Gamma-div-B-eq1}). From the transformational and the dimensional properties listed in Table I, the most general form of $(s'\varphi^\a)$ may be given by 
\bea
\label{eq:divergent-BRST-tr.-B-eq1}
s'V^a 
& = & 
- \displaystyle{\frac{i}{2}} X^a_{\ b} c^b 
+ \displaystyle{\frac{i}{2}}  Y^a_{\ b} {\bar c}^b  \ , \nn 
s'\varpi_a 
& = & 
 \displaystyle{\frac{i}{2}} U_{ab}^{\ c} \varpi_c c^b 
- \displaystyle{\frac{i}{2}} W_{ab}^{\ c} \varpi_c {\bar c}^b  
      \ , \nn 
s'c^a 
& = & 
- \beta \displaystyle{\frac{g}{2}} [c\times c]^a           \ , \quad 
s'{\bar c}^a   
 =  
- \beta \displaystyle{\frac{g}{2}} [{\bar c}\times {\bar c}]^a \ , \nn
s'\phi^a 
& = &  \gamma 2\kappa {\dot c}^a  
+ {\tilde \beta} g [\phi\times c]^a    \ , \nn 
s'{\bar \phi}^a 
& = &  \gamma 2\kappa {\dot {\bar c}}^a  
+ {\tilde \beta} g [{\bar \phi}\times {\bar c}]^a       \ ,  
\ena
where $\beta$, ${\tilde \beta}$ and $\gamma$ are divergent coefficients. The operators of dimension 0, namely $X^a_{\ b}$, $Y^a_{\ b} = X_{\ b}^{a \dagger}$, $U_{ab}^{\ c}$ and $W_{ab}^{\ c} = U_{ab}^{\ c \dagger}$, are arbitrary functions of $V$. By using this definition, (\ref{eq:ST-identity-B-eq2}) yields two conditions. 
One is 
\bea
\label{eq:ST-identity-B-eq3}
s s' + s' s = 0 \ . 
\ena
The other is the condition to determine  
${\tilde \Gamma}_{\rm div}|_{I_\varphi = 0} \equiv {\tilde \Gamma}'_{\rm div} + {\tilde \Gamma}^{\rm FP}_{\rm div}$,       
\bea
\label{eq:ST-identity-B-eq4}
\varint \sum_\a \Big\{ 
\displaystyle{\frac{\delta {\tilde K}|_{I_\varphi = 0}}{\delta \varphi^\a}} (s' \varphi^\a) 
+ \displaystyle{\frac{\delta {\tilde \Gamma}_{\rm div}|_{I_\varphi = 0}}{\delta \varphi^\a}} (s \varphi^\a)
\Big\} = 0 \ , 
\ena
where ${\tilde K}|_{I_\varphi = 0} \equiv K_{\rm BRST}|_{B={\bar B}=0}$. Furthermore, (\ref{eq:ST-identity-B-eq4}) is decomposed into two parts, 
\bea
\label{eq:ST-identity-B-eq5}
& {} & 
\varint \sum_{(V, \varpi, \phi, {\bar \phi})} \Big\{ 
\displaystyle{\frac{\delta K}{\delta \varphi^\a}} (s' \varphi^\a) 
+ \displaystyle{\frac{\delta {\tilde \Gamma}'_{\rm div}}{\delta \varphi^\a}} (s \varphi^\a)
\Big\} = 0 \ ,        \nn 
& {} &
\varint \sum_{(V, \varpi, \phi, {\bar \phi},c,{\bar c})} \Big\{ 
\displaystyle{\frac{\delta K^{\rm FP}}{\delta \varphi^\a}} (s' \varphi^\a) 
+ \displaystyle{\frac{\delta {\tilde \Gamma}_{\rm div}^{\rm FP}}{\delta \varphi^\a}} (s \varphi^\a)
\Big\} = 0 \ , 
\ena
due to the $(c', {\bar c}')$-dependence. 
Here we comment on the generality of (\ref{eq:divergent-BRST-tr.-B-eq1}). 
Since $\phi^a$ and ${\bar \phi}^a$ are dimension-2 quantities, there is a possibility to exchange $\phi$ (or ${\bar \phi}$) and $\varpi$ in the definition (\ref{eq:divergent-BRST-tr.-B-eq1}). In the definition of $s'\phi^a$, the term $g [\varpi \times c]^a$ is excluded because of the chirality of $s'\phi^a$, i.e., $D(s'\phi)= 0$. Because of this chirality, this term cannot be multiplied by an arbitrary function of $V$. In the definition of $s'\varpi^a$, the possible terms might be ${\tilde U}_{ab}^{\ c} c^b \phi^a$ and ${\tilde W}_{ab}^{\ c} {\bar c}^b {\bar \phi}^a$ where ${\tilde U}_{ab}^{\ c}$ and ${\tilde W}_{ab}^{\ c}$ are dimensionless arbitrary functions of $V$. However, from the first condition in (\ref{eq:ST-identity-B-eq5}), since $K$ includes terms linear with respect to $\varpi$, this implies that $\Gamma'_{\rm div}$ must include a $\varpi$-independent term, and this is not the case, as we have explained above. Although this can be checked  directly by adding these terms in the definition of $s'\varpi^a$ given in (\ref{eq:divergent-BRST-tr.-B-eq1}), to simplify the analysis, we discard this possibility in the definition (\ref{eq:divergent-BRST-tr.-B-eq1}). 

We first solve the condition (\ref{eq:ST-identity-B-eq3}). By substituting the definition (\ref{eq:divergent-BRST-tr.-B-eq1}) into (\ref{eq:ST-identity-B-eq3}), we obtain 
\bea
\label{eq:relation-ST-identity-B-eq1}
& {} & 
{\tilde \beta} = \beta \ , \nn 
& {} & 
X^c_{\ [d}\pa_c L^a_{\ b]} + L^c_{[d}\pa_c X^a_{\ b]} 
- igf^{dbc}(\beta L^a_{\ c} + X^a_{\ c}) = 0 \ ,      \nn 
& {} & 
Y^c_{\ [d}\pa_c L^{\ a}_b] + L^{\ c}_{[d}\pa_c Y^a_{\ b]} 
- igf^{dbc}(\beta L^{\ a}_c + Y^a_{\ c}) = 0 \ ,      \nn 
& {} & 
Y^c_{\ d}\pa_cL^a_{\ b} + L_d^{\ c}\pa_c X^a_{\ b} 
- (X^c_{\ b}\pa_c L_d^{\ d} + L^c_{\ b}\pa_c Y^a_{\ d} ) = 0  \ , 
\ena
and 
\bea
\label{eq:relation-ST-identity-B-eq2} 
& {} & 
- L^d_{\ [e}\pa_d U_{ab]}^{\ c} + U_{a[b}^{\ d}\pa_d L^c_{\ e]} 
+ igf^{deb} U_{ad}^{\ c}   \nn 
& {} & \qquad\qquad\qquad\qquad 
- X^d_{\ [e}\pa_d \pa_a L^c_{\ d]} 
+ \pa_a L^d_{\ [b}U_{de]}^{\ c} 
+ ig \beta f^{deb}\pa_a L^c_{\ d} = 0    \ , \nn 
& {} & 
- L^{\ d}_{[e}\pa_d W_{ab]}^{\ c} + W_{a[b}^{\ d}\pa_d L_{e]}^{\ c} 
- igf^{deb} W_{ad}^{\ c}   \nn 
& {} & \qquad\qquad\qquad\qquad 
- Y^d_{\ [e}\pa_d \pa_a L_{b]}^{\ c} 
+ \pa_a L^{\ d}_{[b}W_{de]}^{\ c} 
- ig \beta f^{deb}\pa_a L_d^{\ c} = 0    \ , \nn 
& {} & 
L_e^{\ d}\pa_d U_{ab}^{\ c} - U_{ab}^{\ d}\pa_d L_e^{\ c} 
- (L^d_{\ b}\pa_d W_{ae}^{\ c} - W_{ae}^{\ d} \pa_d L^c_{\ b} )  \nn 
& {} & \qquad\qquad
+ (Y^d_{\ e}\pa_d \pa_a L^c_{\ b} -\pa_a L^d_{\ b} W_{de}^{\ c} ) 
- (X^d_{\ b}\pa_d \pa_a L_e^{\ c} - \pa_a L_e^{\ d} U_{db}^{\ c} ) 
= 0  \ . 
\ena
In (\ref{eq:relation-ST-identity-B-eq1}) and (\ref{eq:relation-ST-identity-B-eq2}), we have used the partial anti-symmetrization defined by $[ab...cd] \equiv {1\over 2}(ab...cd -db...ca)$. 
From (\ref{eq:relation-ST-identity-B-eq1}), the solution of (\ref{eq:relation-ST-identity-B-eq2}) is given by 
\bea
\label{eq:solution-ST-identity-B-eq1}
U_{ab}^{\ c} = \pa_a X^c_{\ b}\ , \ 
W_{ab}^{\ c} = \pa_a Y^c_{\ b}\ . 
\ena
By using the Maurer-Cartan-type equations for $L^a_{\ b}$ and $L_b^{\ a}$ given in (\ref{eq:analogue-Lie-derivative-A-eq4}), we find the solution to (\ref{eq:relation-ST-identity-B-eq1}) 
\bea
\label{eq:solution-ST-identity-B-eq2}
X^a_{\ b} = \a L^a_{\ b} + (\b - \a) V^c \pa_c L^a_{\ b} \ , \quad 
Y^a_{\ b} = \a L_b^{\ a} + (\b - \a) V^c \pa_c L_b^{\ a} \ , 
\ena
where $\a$ is a divergent coefficient. In terms of the $renormalized$ BRST transformation $\delta_{\rm BRST}\varphi^a = \lambda (s\varphi^a)$, (\ref{eq:divergent-BRST-tr.-B-eq1}) with (\ref{eq:relation-ST-identity-B-eq1}) and (\ref{eq:relation-ST-identity-B-eq2}) is expressed as 
\bea
\label{eq:divergent-BRST-tr.-B-eq2}
s'V 
& = & 
\a (sV) + (\b - \a)V^c \pa_c (sV) \ ,  \nn 
s'\varpi 
& = & 
\a (s\varpi) + (\b - \a)\{ (s\varpi) + V^c \pa_c (s\varpi) \}   \ , \nn 
s'c 
& = & 
\beta (sc) \ , \quad 
s'{\bar c} = \beta (s{\bar c})   \ , \nn 
s'\phi 
& = & 
\c (s\phi) + (\b -\c)g[\phi\times c] \ , \quad 
s'{\bar \phi} =  
\c (s{\bar \phi}) + (\b -\c)g[{\bar \phi}\times {\bar c}]  \ . 
\ena
Next, we solve the condition (\ref{eq:ST-identity-B-eq4}) and determine ${\tilde \Gamma}_{\rm div}$. By substituting (\ref{eq:divergent-BRST-tr.-B-eq2}) into (\ref{eq:ST-identity-B-eq5}), it becomes 
\bea
\label{eq:tilde-Gamma-div-ST-identity-B-eq1} 
& {} & 
\int_0 \!\!\! \Big\{ 
(\b-\a)\Big( 
\displaystyle{\frac{\delta K}{\delta V}} (s''V) 
+ \displaystyle{\frac{\delta K}{\delta \varpi}} (s\varpi + s''\varpi) 
\Big) 
+ \displaystyle{\frac{\delta {\tilde \Gamma}'_{\rm div}}{\delta V}} (sV)
+ \displaystyle{\frac{\delta {\tilde \Gamma}'_{\rm div}}{\delta \varpi}} 
 (s\varpi)
\Big\}          \nn 
& {} & 
+ \int_{+} \!\!\! \Big\{ 
(\c-\a) \displaystyle{\frac{\delta K}{\delta \phi}} (2\kappa {\dot c}) 
+ (\b -\a) \displaystyle{\frac{\delta K}{\delta \phi}} g[\phi\times c] 
+ \displaystyle{\frac{\delta {\tilde \Gamma}'_{\rm div}}{\delta \phi}} (s\phi) 
\Big\}         \nn 
& {} & 
+ \int_{-} \!\!\! \Big\{ 
(\c-\a) \displaystyle{\frac{\delta K}{\delta {\bar \phi}}} (2\kappa {\dot {\bar c}}) 
+ (\b -\a) \displaystyle{\frac{\delta K}{\delta {\bar \phi}}} g[{\bar \phi}\times {\bar c}] 
+ \displaystyle{\frac{\delta {\tilde \Gamma}'_{\rm div}}{\delta {\bar \phi}}} (s{\bar \phi}) 
\Big\} = 0   \ ,
\ena
where $s''V^a \equiv V^c \pa_c (sV^a)$ and $s''\varpi^a \equiv V^c \pa_c (s\varpi^a)$. The general solution of this equation consists of two parts. One is the non-trivial cohomology of the BRST transformation, i.e., the solution of 
$\sum_{\a \in (V, \varpi, \phi, {\bar \phi})} \varint (s\varphi^a) ( \delta{\tilde \Gamma}'_{\rm div}/ \delta \varphi^\a ) = 0$. The other is a particular solution of (\ref{eq:tilde-Gamma-div-ST-identity-B-eq1}). 

A particular solution of (\ref{eq:tilde-Gamma-div-ST-identity-B-eq1}) is given by 
\bea
\label{eq:particular-solution-B-eq1}
& {} & \!\!\!\!\!\!\!
(\b-\a)\Big\{ 
\int_0 \!\!\! V^a \displaystyle{\frac{\delta K}{\delta V^a}} 
+ \int_{+} \!\!\! \phi^a \displaystyle{\frac{\delta K}{\delta \phi^a}} 
+ \int_{-} \!\!\! {\bar \phi}^a \displaystyle{\frac{\delta K}{\delta {\bar \phi}^a}}  \Big\}    \nn
& {} & \qquad\qquad\qquad\qquad\qquad\qquad
+ 
(\a-\c)\Big\{ 
 \int_{+} \!\!\! \phi^a \displaystyle{\frac{\delta K}{\delta \phi^a}} 
+ \int_{-} \!\!\! {\bar \phi}^a \displaystyle{\frac{\delta K}{\delta {\bar \phi}^a}} 
\Big\}   \  . 
\ena
This can be directly checked by using the commutation relations 
\bea
\label{eq:commutation-relations-B-eq1}
\Big[ \int_0 \!\!\! V^a \displaystyle{\frac{\delta}{\delta V^a}} ,\  \delta'_{\rm BRST} \Big] 
& = & 
\int_0 \!\!\! \Big(  
(s''V^a)\displaystyle{\frac{\delta}{\delta V^a}} 
+ (s''\varpi^a) \displaystyle{\frac{\delta}{\delta \varpi^a}} 
- (sV^a)\displaystyle{\frac{\delta}{\delta V^a}} 
\Big) \ ,    \nn 
{\Big[} \int_0 \!\!\! \varpi^a \displaystyle{\frac{\delta}{\delta \varpi^a}} ,\  \delta'_{\rm BRST} \Big]   
& = &  0 \ ,    \nn 
{\Big[} \int_{+} \!\!\! \phi^a \displaystyle{\frac{\delta}{\delta \phi^a}} ,\  \delta'_{\rm BRST} \Big] 
& = & 
\int_{+} \!\!\! \Big( g[\phi \times c ]\displaystyle{\frac{\delta}{\delta \phi^a}}  
- (s\phi^a)\displaystyle{\frac{\delta}{\delta \phi^a}}  \Big) 
         \  , \nn
{\Big[} \int_{-} \!\!\! {\bar \phi}^a \displaystyle{\frac{\delta}{\delta {\bar \phi}^a}} ,\  \delta'_{\rm BRST} \Big] 
& = & 
\int_{-} \!\!\! \Big( g[{\bar \phi} \times {\bar c}]\displaystyle{\frac{\delta}{\delta {\bar \phi}^a}}    
- (s{\bar \phi}^a) \displaystyle{\frac{\delta}{\delta {\bar \phi}^a}} \Big)             \  ,  
\ena
where 
$\delta'_{\rm BRST} 
= \sum_{\a \in (V, \varpi, \phi, {\bar \phi})} \varint (s\varphi^\a) {\delta/\delta \varphi^\a} $. 
For the non-trivial cohomology of the BRST transformation, there exist three types of possible forms which are quadratic and linear with respect to $\varpi$. We assume their most general forms, 
\bea
\label{eq:non-trivial-cohomology-B-eq1}
\int_0 \!\!\! \Sigma^{ab}\varpi^a \varpi^b \ ,  \ 
\int_0 \!\!\! \varpi_a \Xi^a \ , \ 
\int_0 \!\!\! \varpi_a ( 
\Omega^a_{\ b} {\dot V}^b + {\tilde \Omega}^a_{\ b} \phi^b 
+ {\tilde \Omega}^{a \dagger}_{\ b} {\bar \phi}^b ) \ .  
\ena
Here $\Sigma^{ab}$, $\Xi^a$, $\Omega^a_{\ b}$ and ${\tilde \Omega}^a_{\ b}$, which are arbitrary functions of $V$, have dimensions 0, 2, 0 and 0, respectively.  It is not difficult to show that 
$\Sigma^{ab} \propto G^{ab}$, 
$\Omega^a_{\ b} \propto \delta^a_{\ b}$, 
${\tilde \Omega}^a_{\ b} \propto L^a_{\ b}$ 
and 
${\tilde \Omega}^{a\ \dagger}_{\ b} \propto L_b^{\ a}$. 
In particular, in the third invariant, the relative numerical factors are fixed by the BRST invariance. For $\Xi$ in the second invariant in (\ref{eq:non-trivial-cohomology-B-eq1}), we obtain 
\bea
\label{eq:non-trivial-cohomology-B-eq2}
(\pa_c L^a_{\ b}\Xi^c - L^c_{\ b}\pa_c \Xi^a )c^b 
-( \pa_c L_b^{\ a} \Xi^c - L_b^{\ c}\pa_c \Xi^a ){\bar c}^b = 0 \ .
\ena
In order to clarify the meaning of this condition, we redefine 
$\Xi^a$ as 
$\Xi^a \equiv L^a_{\ b}{\tilde \Xi}^b_1$ 
and 
$\Xi^a \equiv L_b^{\ a}{\tilde \Xi}^b_2$. 
Then (\ref{eq:non-trivial-cohomology-B-eq2}) reads 
\bea
\label{eq:non-trivial-cohomology-B-eq3}
c^b {\cal E}_b^{({\rm R})}{\tilde \Xi}_1 
= - [\ c, {\tilde \Xi}_1\ ] \ , \ 
{\bar c}^b  {\cal E}_b^{({\rm L})}{\tilde \Xi}_1 
= 0                \ , 
\ena
and
\bea
\label{eq:non-trivial-cohomology-B-eq4}
c^b {\cal E}_b^{({\rm R})}{\tilde \Xi}_2 
= 0 \ , \ 
{\bar c}^b  {\cal E}_b^{({\rm L})}{\tilde \Xi}_2 
=  [\ {\bar c}, {\tilde \Xi}_2\ ]      \ ,  
\ena
where 
${\cal E}_a^{({\rm R})} \equiv {1\over 2g}L^b_{\ a}\pa_b$,  
${\cal E}_a^{({\rm L})} \equiv {1\over 2g}L^{\ b}_a \pa_b$ 
and  
${\tilde \Xi}_{1(2)} \equiv {\tilde \Xi}_{1(2)}^a t^a$.  
Now, (\ref{eq:non-trivial-cohomology-B-eq3}) and (\ref{eq:non-trivial-cohomology-B-eq4}) simply mean that ${\tilde \Xi}_1$ and ${\tilde \Xi}_2$ are transformed as 
${\tilde \Xi}_1 \rightarrow e^{ig\lambda c} {\tilde \Xi}_1 e^{-ig\lambda c}$ 
and 
${\tilde \Xi}_2 \rightarrow e^{ig\lambda {\bar c}} {\tilde \Xi}_2 e^{-ig\lambda {\bar c}}$ 
under the BRST transformation of the vector superfield 
$e^{2gV} \rightarrow 
e^{ig\lambda{\bar c}} e^{2gV} e^{-ig\lambda c}$.
Such operators of dimension $3/2$ are the so-called gluino fields, 
$W_\a$ and ${\bar W}_{\dot \a}$. To construct the operators of dimension 2, one more local gauge covariant spinor derivative, ${\cal D}_\a \equiv e^{-2gV}D_\a e^{2gV}$ or ${\bar {\cal D}}_{\dot \a} \equiv e^{2gV} {\bar D}_{\dot \a} e^{-2gV}$, is necessary. Therefore, the second invariant is given by 
${\tilde \Xi}_1 \propto \{ {\cal D}^\a,\ W_\a \}$ and ${\tilde \Xi}_2 \propto \{ {\bar {\cal D}}_{\dot \a},\ {\bar W}^{\dot \a}\}$. 
Because of the reality condition, these two terms are not independent. 
The uniqueness of the operators ${\tilde \Xi}_1$ and ${\tilde \Xi}_2$ of dimension 2 is confirmed as follows. The operators of dimension 2, which do not depend on the stochastic-time derivative, are constructed from $e^{2gV}$, $e^{-2gV}$, $D_\a$ and ${\bar D}_{\dot \a}$. They must include four spinor derivatives. For the transformation property, they also must consist of the same number of $e^{2gV}$ and $e^{-2gV}$. In particular, because of the chirality of the FP ghosts $c$ and ${\bar c}$, the spinor derivatives that exhibit the proper transformation property are ${\cal D}_\a$ and ${\bar D}_{\dot \a}$ for ${\tilde \Xi}_1$. Therefore, possible forms of ${\tilde \Xi}_1$ can be constructed by successive operations of these derivatives on ${\cal O}$, which is a quantity transformed as $
{\cal O} \rightarrow 
e^{ig\lambda c} {\cal O} e^{-ig\lambda c}$. Since the commutation relations of four spinor derivatives, using ${\cal D}_\a$ and ${\bar D}_{\dot \a}$, are expressed by $\{ {\cal D}^\a, W_\a \}$, $W^\a {\cal D}_\a$ and ${\cal D}^m {\cal D}_m$, where 
$\{ {\cal D}_\a,\ {\bar D}_{\dot \a} \} = -2i \sigma^m_{\a {\dot \a}} {\cal D}_m$, it is found that $\{ {\cal D}^\a,\ W_\a \}$ is the unique possibility for ${\tilde \Xi}_1$. 
We can also conclude that $\{ {\bar {\cal D}}_{\dot \a},\ {\bar W}^{\dot \a} \}$ is the unique possibility for ${\tilde \Xi}_2$.

In this way, we obtain the general solution of the reduced form of the Ward-Takahashi identity (\ref{eq:tilde-Gamma-div-ST-identity-B-eq1})
\bea
\label{eq:solution-ST-identity-B-eq3}
{\tilde \Gamma}'_{\rm div} 
& = & 
a'K^{(1)} + b'K^{(2)} + c'K^{(3)} 
+ (\b-\a)\Big\{ 
\int_0 \!\!\! V^a \displaystyle{\frac{\delta K}{\delta V^a}} 
+ \int_{+} \!\!\! \phi^a \displaystyle{\frac{\delta K}{\delta \phi^a}} 
+ \int_{-} \!\!\! {\bar \phi}^a \displaystyle{\frac{\delta K}{\delta {\bar \phi}^a}}  \Big\}    \nn 
& {} & \qquad\qquad 
+ 
(\a-\c)\Big\{ 
 \int_{+} \!\!\! \phi^a \displaystyle{\frac{\delta K}{\delta \phi^a}} 
+ \int_{-} \!\!\! {\bar \phi}^a \displaystyle{\frac{\delta K}{\delta {\bar \phi}^a}} 
\Big\} 
\ena
Here, $K^{(1)}$, $K^{(2)}$ and $K^{(3)}$ are defined by 
\bea
\label{eq:solution-ST-identity-eq4}
K^{(1)} 
& = & 
\int_0 \!\! \displaystyle{\frac{1}{2\kappa}} G^{ab}\varpi_a \varpi_b  \ , \nn 
K^{(2)} 
& = & 
- i \int_0 \!\! \displaystyle{\frac{1}{4\kappa g}}
\varpi_a \Big( 
L^a_{\ b}\{ {\cal D}^\a, W_\a \} + L_b^{\ a}\{ {\bar {\cal D}}_{\dot \a}, {\bar W}^{\dot \a}  \}    
\Big)    \ , \nn
K^{(3)}
& = & 
\int_0 \!\! \varpi_a \Big\{ 
i {\dot V}^\a - \displaystyle{\frac{1}{4\kappa}} (L^a_{\ b}\phi^b - {\bar \phi}^b L_b^{\ a}) 
\Big\}      \ . 
\ena
The divergent terms in the effective stochastic action are given by 
${\tilde \Gamma}_{\rm div} 
= {\tilde \Gamma}'_{\rm div} + {\tilde \Gamma}_{\rm div}^{({\rm FP})} + {\tilde \Gamma}_{\rm div}^{({\rm ex})}$, where the other two terms are defined in (\ref{eq:tilde-Gamma-div-B-eq1}). The structure of the divergences ${\tilde \Gamma}_{\rm div}$ is similar to the tree-level stochastic action, and thus we finally show that the divergences are absorbed by the renormalization of the wave functions and the coupling constants.   

Let us consider the scale transformation
\bea
\label{eq:scale-tr.-B-eq1}
V 
& \rightarrow & 
e^{x \rho} V,\ 
\varpi \rightarrow e^{y \rho} \varpi,\ 
\phi \rightarrow e^{z \rho} \phi,\ 
{\bar \phi} \rightarrow e^{z \rho} {\bar \phi}, \nn
g 
& \rightarrow & 
e^{-x \rho}g ,\ 
\kappa 
\rightarrow e^{(z-x)} \kappa \ .
\ena
Under the condition 
\bea
\label{eq:scale-tr.-B-eq2}
x=\displaystyle{\frac{1}{2}} (-a'+ b'+ c')\ , \ 
y=\displaystyle{\frac{1}{2}} (a'-b'+ c')\ , \ 
z=\displaystyle{\frac{1}{2}} (-a'- b'+ 3c')\ , \ 
\ena
we have 
\bea
\label{eq:scale-tr.-B-eq3}
& {} & 
a' K^{(1)} + b' K^{(2)} + c' K^{(3)}        
=  
\Big\{ 
\int_0 \!\! \Bigl( x V^a \displaystyle{\frac{\delta}{\delta V^a}}    
+ y\varpi^a \displaystyle{\frac{\delta}{\delta \varpi^a}} \Bigr)      \nn 
& {} & \qquad\qquad 
+ z\Bigl( \int_{+} \!\!\! \phi^a \displaystyle{\frac{\delta}{\delta \phi^a}}  
+ \int_{-} \!\!\! {\bar \phi}^a \displaystyle{\frac{\delta}{\delta {\bar \phi}^a}} \Bigr)  - x g \displaystyle{\frac{\pa}{\pa g}} + (z-x) \kappa \displaystyle{\frac{\pa}{\pa \kappa}} \Big\}K    \ . 
\ena
Then, using this relation, ${\tilde \Gamma}'_{\rm div}$ can be written as 
\bea
\label{eq:renormalization-B-eq1}
{\tilde \Gamma}'_{\rm div} 
& = & 
\Big\{ 
(x+\b-\a)\int_0 \!\! V^a \displaystyle{\frac{\delta}{\delta V^a}} 
+ y\int_0 \!\! \varpi^a \displaystyle{\frac{\delta}{\delta \varpi^a}}  \nn 
& {} & \quad 
+ (z+\b-\c)\Bigl( \int_{+} \!\!\! \phi^a \displaystyle{\frac{\delta}{\delta \phi^a}}  + \int_{-} \!\!\! {\bar \phi}^a \displaystyle{\frac{\delta}{\delta {\bar \phi}^a}} \Bigr)  - x g \displaystyle{\frac{\pa}{\pa g}} + (z-x) \kappa \displaystyle{\frac{\pa}{\pa \kappa}} \Big\}K    \ . 
\ena
This expression can be expressed in terms of 
$K_{\rm tot} \equiv K_{\rm BRST} + K_{\rm ex}^{\rm BRST}$. After some simple algebraic manipulations, we obtain 
\bea
\label{eq:renormalization-B-eq2}
& {} & {\tilde \Gamma}_{\rm div} \equiv {\tilde \Gamma}'_{\rm div} + {\tilde \Gamma}^{({\rm FP})}_{\rm div} + {\tilde \Gamma}^{({\rm ex})}_{\rm div}   \nn 
& {} & 
= 
\Big[
(x+\b-\a)\int_0 \!\! \Big( V^a \displaystyle{\frac{\delta}{\delta V^a}} 
- I_{V^a} \displaystyle{\frac{\delta}{\delta I_{V^a}}} \Big)
+ y\int_0 \!\! \Big( \varpi^a \displaystyle{\frac{\delta}{\delta \varpi^a}} 
- I_{\varpi^a} \displaystyle{\frac{\delta}{\delta I_{\varpi^a}}} \Big) \nn 
& {} & \quad 
+ (z+\b-\c)\Big\{ 
\int_{+} \!\!\! \Big( \phi^a \displaystyle{\frac{\delta}{\delta \phi^a}} 
- I_{\phi^a} \displaystyle{\frac{\delta}{\delta I_{\phi^a}}} \Big) 
+ \int_{-} \!\!\! \Big( {\bar \phi}^a \displaystyle{\frac{\delta}{\delta {\bar \phi}^a}}  
- I_{{\bar \phi}^a} \displaystyle{\frac{\delta}{\delta I_{{\bar \phi}^a}}}      \Big)     \Big\}          \nn 
& {} & \quad 
+ (\b+x)\Big\{ 
\int_{+} \!\!\! \Big( c^a \displaystyle{\frac{\delta}{\delta c^a}} 
- I_{c^a} \displaystyle{\frac{\delta}{\delta I_{c^a}}} \Big) 
+ \int_{-} \!\!\! \Big( {\bar c}^a \displaystyle{\frac{\delta}{\delta {\bar c}^a}}  
- I_{{\bar c}^a} \displaystyle{\frac{\delta}{\delta I_{{\bar c}^a}}}      \Big)     \Big\}          \nn 
& {} & \quad 
+ (\c-\b-x)\Big\{ \int_{+} \!\!\! \Big( B^a \displaystyle{\frac{\delta}{\delta B^a}} 
- I_{c^{'a}} \displaystyle{\frac{\delta}{\delta I_{c^{'a}}}} \Big) 
+ \int_{-} \!\!\! \Big( {\bar B}^a \displaystyle{\frac{\delta}{\delta {\bar B}^a}}   
- I_{{\bar c}^{'a}} \displaystyle{\frac{\delta}{\delta I_{{\bar c}^{'a}}}} \Big)      \Big\}         \nn 
& {} & \quad 
+ (\c-\b-x)\Big( \int_{+} \!\!\!  c^{'a} \displaystyle{\frac{\delta}{\delta c^{'a}}} + \int_{-} \!\!\! {\bar c}^{'a} \displaystyle{\frac{\delta}{\delta {\bar c}^{'a}}} \Big)                     \nn
& {} & \quad 
- x g \displaystyle{\frac{\pa}{\pa g}} + (z-x) \kappa \displaystyle{\frac{\pa}{\pa \kappa}} + (\a -\c + z - x) \xi \displaystyle{\frac{\pa}{\pa \xi}}
\Big] K_{\rm tot}    \ . 
\ena
%
%
%
This equation completely fixes the renormalization constants, which are introduced as 
\bea
\varphi_0^\a = Z^{1/2}_{\varphi^\a} \varphi^\a ,\ 
I_{\varphi^\a 0} = Z^{1/2}_{I \varphi} I_{\varphi^\a} ,\ 
g_0 = Z_g g ,\ \kappa_0 = Z_\kappa \kappa ,\ \xi_0 = Z_\xi \xi , 
\ena
where 
$\varphi^\a = 
(V, \varphi, c, {\bar c}, c', {\bar c}', \phi, {\bar \phi}, B, {\bar B})$. 
By the definitions of the renormalization constants, (\ref{eq:renormalization-B-eq2}) can be written 
\bea
\label{eq:list-renormalization-constant-B-eq1} 
& {} & 
K_{\rm tot} (\varphi^\a_0, I_{\varphi^\a 0}, g_0, \kappa_0, \xi_0) 
\equiv K_{\rm tot} (\varphi^\a, I_{\varphi^\a}, g, \kappa, \xi) - \Gamma^{(1)}_{\rm div} + {\cal O} (\hbar^2)          , \nn 
& {} &
1 - Z_V^{1/2} = x + \b - \a \ , \ 1 - Z_\varpi^{1/2} = y \ , \
1 - Z_\phi^{1/2} = 1 - Z_{\bar \phi}^{1/2} = z + \b - \c    \ , \nn
& {} & 
1 - Z_{c'}^{1/2} = 1 - Z_{{\bar c}'}^{1/2} = \c - \b - x 
= 
1 - Z_B^{1/2} = 1 - Z_{\bar B}^{1/2}\ , \nn
& {} & 
1 - Z_{c}^{1/2} = 1 - Z_{\bar c}^{1/2} = \b + x \ , \ 
1 - Z_\xi^{1/2} = \a - \c + z - x \ ,     \nn 
& {} & 
Z_{\varphi^\a} Z_{I \varphi^\a} = 1,\  {\rm for}\ (V, \varpi, c, {\bar c}, c', {\bar c}', \phi, {\bar \phi})    \ . 
\ena
In particular, we note that the relation $Z_{\varphi^\a} Z_{I \varphi^\a} = 1$ ensures that we can renormalize the divergences of the effective stochastic action within this multiplicative renormalization procedure in the arbitrary order of the perturbation. 

The independent divergent coefficients are $(x,y,z) \sim (a',b',c')$ and $(\alpha,\beta,\gamma)$, while there are nine renormalization constants. This implies that there are three relations for the renormalization constants. In fact, we have 
\bea
\label{eq:}
Z_\kappa^{-1} Z_B^{1/2} Z_\phi^{1/2} = 1,\ 
Z_\phi^{1/2} = Z_\xi Z_V^{1/2} = Z_{\bar \phi}^{1/2} ,\ 
Z_B = Z_{\bar B} = Z_{c'}=Z_{{\bar c}'}  \ . 
\ena
The first two relations are a consequence of the non-renormalization of the stochastic gauge fixing term. The last one implies that the BRST transformation 
$sc' = iB$ ($s{\bar c}' = i{\bar B}$) is not renormalized. 

The analysis given in this appendix completes the proof of the renormalizability of the SQM approach in terms of the BRST invariant stochastic action principle in a formal sense. In the actual calculation, since the interaction is non-polynomial in the superfield formalism, we need to apply the background field method (BFM). We note that, in BFM, the relations for the renormalization constants may be modified, for example, 
$Z_g Z_V^{1/2} = 1$ holds in BFM.    

Although, we have not given a similar proof for the truncated version, (\ref{eq:stochastic-BRST-tr.-eq2}) and (\ref{eq:BRST-invariant-stochastic-action-eq2}), 
it could be possible to provide such a proof in terms of the Ward-Takahashi identity, which is also given by a form similar to (\ref{eq:ST-identity-B-eq1}). 
\bea
\label{eq:truncated-ST-identity-B-eq1}
\varint \sum_\a \displaystyle{\frac{\delta \Gamma}{\delta \varphi^\a}} 
\displaystyle{\frac{\delta \Gamma}{\delta I_{\varphi^\a}}} 
= \Gamma \ast \Gamma = 0        \ , 
\ena
where $\varphi^\a = (V, \varpi, c, {\bar c}, c', {\bar c}')$. The external source $I_{\varphi^\a}$ is coupled to $s\varphi^\a$, which is defined by the truncated BRST transformation (\ref{eq:stochastic-BRST-tr.-eq2}). 
In this case, the local identities for the connected stochastic Feynman diagrams, which are derived by the change of the integration variables ($c' \rightarrow c' + \delta c'$ and ${\bar c}' \rightarrow {\bar c}' + \delta{\bar c}'$), cannot be converted to those for the effective stochastic action. This indicates that the stochastic gauge fixing terms may be renormalized in the truncated version (\ref{eq:BRST-invariant-stochastic-action-eq2}). If this is indeed the case, it would complicate the standard argument for the renormalizability. In particular, for the part of the stochastic gauge fixing and the F-P ghost sector, we would have to treat the most general form of the gauge fixing, because the truncated BRST symmetry is independent of the specified form of the stochastic gauge fixing functions. For example, in the renormalization procedure of the truncated version (\ref{eq:BRST-invariant-stochastic-action-eq2}), it seems that there is no restriction on the modification of the gauge fixing functions, 
%
$\Phi = \phi - {i\over4}\xi{\bar D}^2D^2 h(V)$   
and 
${\bar \Phi} = {\bar \phi} + {i\over4}\xi D^2{\bar D}^2 h(V)$, 
where $h(V)$ is an arbitrary function of $V$. However, we emphasize that the BRST symmetry ensures that the expectation values of the local gauge invariant observables are $independent$ of the stochastic gauge fixing functions and that the FP ghosts do $not$ contribute to the renormalization.


\end{document}